\documentclass[graybox, nosecnum]{svmult}

\usepackage{mathptmx}       
\usepackage{helvet}         
\usepackage{courier}        
\usepackage{type1cm}        
\usepackage{makeidx}         
\usepackage{graphicx}        
\usepackage{cite}
\usepackage{dcolumn}
\usepackage{bm}
\usepackage{epstopdf}
\usepackage{float}
\usepackage{xcolor, colortbl}
\usepackage{multicol}        
\usepackage[bottom]{footmisc}
\usepackage{hyperref}        
\usepackage{soul}            
\usepackage{CJK}
\usepackage{amsmath}        
\usepackage{bm}
\usepackage{animate}
\usepackage{subfigure}

\makeindex             

\begin{document}
\title*{Model for collective motion}
\author{Z. P. Li \thanks{zpliphy@swu.edu.cn} and D. Vretenar}
\institute{Z. P. Li \at School of Physical Science and Technology, Southwest University, Chongqing 400715, China, \email{zpliphy@swu.edu.cn}
\and D. Vretenar \at Department of Physics, Faculty of Science, University of Zagreb, HR-10000 Zagreb, Croatia;
State Key Laboratory of Nuclear Physics and Technology,
School of Physics, Peking University, Beijing 100871, China, \email{vretenar@phy.hr}}
%
%
\maketitle
\abstract{Collective motion is a manifestation of emergent phenomena in medium-heavy and heavy nuclei. A relatively large number of constituent nucleons contribute coherently to nuclear excitations (vibrations, rotations) that are characterized by large electromagnetic moments and transition rates. 
Basic features of collective excitations are reviewed, and a simple model introduced that describes  large-amplitude quadrupole and octupole shape dynamics, as well as the dynamics of induced fission. Modern implementations of the collective Hamiltonian model are based on the microscopic framework of energy density functionals, that provide an accurate global description of nuclear ground states and collective excitations. Results of illustrative calculations are discussed in comparison with available data.}

\section{\textit{Outline}}
\vspace{-0.5cm}
\begin{itemize}
\hypersetup{hidelinks}
\item \hyperref[sec:Introduction]{Introduction} 
\item \hyperref[sec:Parameters]{Nuclear Shape Parameters}
\item \hyperref[sec:SO]{Nuclear Surface Oscillations}
\item \hyperref[sec:RVM]{The Rotation-Vibration Model}
\item \hyperref[sec:GCM+GOA]{Microscopic Derivation of the Collective Hamiltonian}
\item \hyperref[sec:MCH-CDFT]{Microscopic Collective Hamiltonian Based on Density Functional Theory}
\end{itemize}

\newpage
\section{\label{sec:Introduction}Introduction}
A medium-heavy or heavy atomic nucleus presents a typical example of a complex quantum system, in which different interactions between a relatively large number of constituent nucleons give rise to physical phenomena that are qualitatively different from those exhibited by few-nucleon systems. There are a number of features that characterize complex systems, but for the topic of the present chapter of particular interest is the emergence of collective structures and dynamics that do not occur in light nuclei composed of only a small number of nucleons. 

Collective motion is the simplest manifestation of emergent phenomena in atomic nuclei. It can be interpreted as a kind of motion in which a large number of nucleons contribute coherently to produce a large amplitude oscillation of one or more electromagnetic multipole moments. Collective motion gives rise to excited states that are characterized by large electromagnetic transition rates to the ground state, that is, rates that correspond to many single-particle transitions \cite{Rowe1970}. This chapter will mainly focus on low-energy large-amplitude collective motion (LACM), such as surface vibrations, rotations, and fission. Theoretical studies of LACM started as early as in the 1930s. Based on the liquid drop model of the atomic nucleus \cite{Weizsacker1935}, Fl\"{u}gge applied Rayleigh's normal modes \cite{Rayleigh1879} to a classical description of low-energy excitations of spherical nuclei \cite{Flugge1941}. The model of Fl\"{u}gge was quantized by Bohr \cite{Bohr1952}, who formulated a  quantum model of surface oscillations of spherical nuclei, and also introduced the concept of intrinsic frame of reference for a quadrupole deformed nuclear surface characterized by the Euler angles and the shape parameters $\beta$ and $\gamma$ (nowadays often called the Bohr deformation parameters).  Subsequently, Bohr and Mottelson \cite{Bohr1953} generalized the model to vibrations and rotations of deformed nuclei. A generalization of the Bohr Hamiltonian to describe large-amplitude collective quadrupole excitations of even-even nuclei of arbitrary shape was introduced by Belyaev \cite{Belyaev1965} and Kumar and Baranger \cite{Kumar1967}.  Several specific forms of the collective Hamiltonian, designed to describe collective excitations in nuclei of particular shape were also  considered \cite{Wilets1956,Davydov1958,Davydov1960,Greiner1962a,Greiner1962b,Faessler1965}.

In the past several decades, enormous progress has been made in developing microscopic many-body theories of nuclear systems. However, a description of collective phenomena starting from single-nucleon degrees of freedom still presents a considerable challenge. One of the methods that has been used to obtain such a description is the adiabatic time-dependent HFB theory (ATDHFB) \cite{Belyaev1965,Baranger1968,Baranger1978} which, in the case of quadrupole collective coordinates, leads to the Bohr Hamiltonian. Another approach to collective phenomena that is based on microscopic degrees of freedom, is the generator coordinate method (GCM) \cite{Hill1953}. In the Gaussian overlap approximation (GOA) \cite{Brink1968,Onishi1975}, the GCM also leads to the Bohr collective Hamiltonian \cite{Une1976,Gozdz1985}.



\section{\label{sec:Parameters}Nuclear Shape Parameters}

Excitation spectra of even-even nuclei in the energy range of up to $\approx 3$ MeV, exhibit characteristic band structures that are interpreted as vibrations and rotations of the nuclear surface in the geometric collective model, first introduced by Bohr and Mottelson \cite{Bohr1952,Bohr1953}, and further elaborated by Faessler and Greiner \cite{Greiner1962a,Greiner1962b}.

For low-energy excitations the compression mode is not relevant because of high incompressibility of nuclear matter, and the diffuseness of the nuclear surface layer can also be neglected to a good approximation. One therefore starts with the model of a nuclear liquid drop of constant density and sharp surface. With these assumptions, the time-dependent nuclear surface can, quite generally, be described by an expansion in spherical harmonics with shape parameters as coefficients:
\begin{equation}
\label{eq:surface}
R(\theta, \phi; t)=R_{0}\left(1+\sum_{\lambda=0}^{\infty} \sum_{\mu=-\lambda}^{\lambda} \alpha_{\lambda \mu}(t)^* Y_{\lambda \mu}(\theta, \phi)\right)
\end{equation}
where $R(\theta, \phi; t)$ denotes the nuclear radius in spherical coordinates $(\theta, \phi)$, and $R_{0}$ is the radius of a sphere with the same volume. The shape parameters $\alpha_{\lambda \mu}(t)$ play the role of collective dynamical variables, and their physical meaning will be discussed for increasing values of $\lambda$.
\vspace{-0.5cm}
\begin{figure}
	\centering
	\includegraphics[width=1\linewidth]{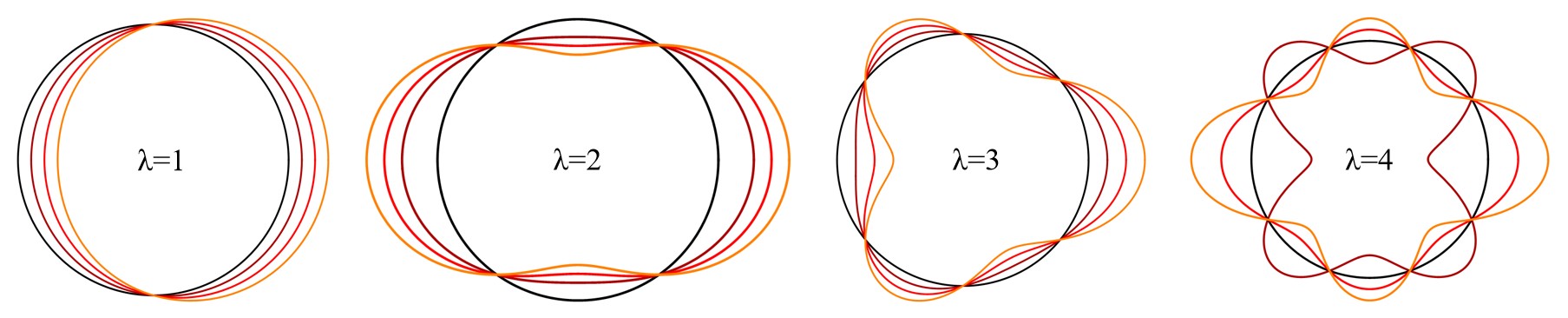}
	\caption{Nuclear shapes with dipole ($\lambda=1$), quadrupole ($\lambda=2$), octupole ($\lambda=3$), and hexadecupole ($\lambda=4$) deformations.}
	\label{fig:shapes}
\end{figure}

To lowest order, the dipole mode $\lambda = 1$ corresponds to a translation of the nucleus as a whole and, therefore, is not considered for low-energy excitations. Dynamical quadrupole deformations, that is, the mode with $\lambda = 2$, turn out to be the most relevant low-lying collective excitations. Most of the following discussion of collective models will focus on this case, so a more detailed description is included below.

Octupole dynamical deformations, $\lambda = 3$, are the principal asymmetric modes of a nucleus associated with negative-parity bands. While there is no evidence for pure hexadecupole excitations in low-energy spectra, this mode plays an important role as admixture to quadrupole excitations, and for fission dynamics. Shape oscillations of higher multipoles are not relevant for  low-energy excitations.

For the case of pure quadrupole deformation the nuclear surface is parameterized 
\begin{equation}
\label{eq:quadef}
R(\theta, \phi)=R_{0}\left(1+\sum_{\mu=-2}^{2} \alpha_{\mu}^* Y_{2 \mu}(\theta, \phi)\right)
\end{equation}
where the time dependence is implicit for dynamical variables. 
If the shape of the nucleus is an ellipsoid, its three principal axes ($x$, $y$, $z$) are linked with the ($X$, $Y$, $Z$) axes of a Cartesian coordinate system in the laboratory frame.
From the symmetry of the ellipsoid, it follows that $a_{1}=a_{-1}=0, a_{2}=a_{-2}$, where $a_\mu$ are the shape parameters in the principal-axis system. Obviously, the five coefficients $\alpha_{\mu}$ in the laboratory frame reduce to two real independent variables $a_0$ and $a_2$ in the principal-axis system, which, together with the three Euler angles, provide a complete parameterization of the nuclear surface. The details of the transformation between $\alpha_{\mu}$ and $a_{\mu}$ are included below.
In the principal-axis system, the nuclear radius is given by
\begin{equation}
\label{eq:quad1}
R(\theta^\prime, \phi^\prime)=R_{0}\left[1+a_{0} Y_{20}(\theta^\prime, \phi^\prime)+a_{2} Y_{22}(\theta^\prime, \phi^\prime)+a_{2} Y_{2-2}(\theta^\prime, \phi^\prime)\right] .
\end{equation}
Two parameters $(a_0,\ a_2)$ are generally used to describe quadrupole deformations but, instead of $a_0$  and  $a_2$, the polar coordinates $\beta$ and $\gamma$ are usually employed. They are defined as follows:
\begin{equation}
\begin{aligned}
\label{eq:polarco}
a_{0}&=\beta \cos \gamma, \\
a_{2}&=\frac{1}{\sqrt{2}} \beta \sin \gamma.
\end{aligned}
\end{equation}
Using Eq.~(\ref{eq:quad1}) and Eq.~(\ref{eq:polarco}), we can express the increments of the three semi-axes in the principal-axis system:
\begin{equation}
\label{eq:delR}
\delta R_{\kappa}=R_{0} \sqrt{\frac{5}{4 \pi}} \beta \cos \left(\gamma-\frac{2 \pi}{3} \kappa\right), \quad \kappa=1,2,3
\end{equation}
where $\kappa=1,2,3$ correspond to $x,y,z$, respectively. 
The parameters $\beta$ and $\gamma$ only describe exactly ellipsoidal shapes in the limit of small $\beta$-values.

\begin{figure}[ht]
	\centering
	\subfigure[]{%
	\includegraphics[width=6.7cm]{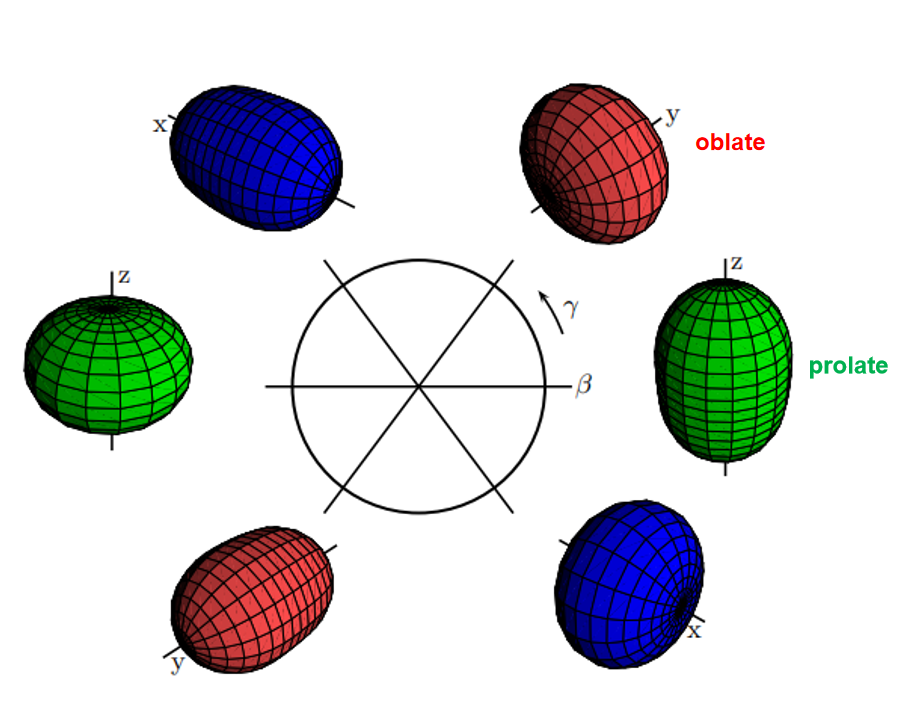}
	\label{fig:quadshapes1}}
	\quad
	\subfigure[]{%
		\includegraphics[width=4.1cm]{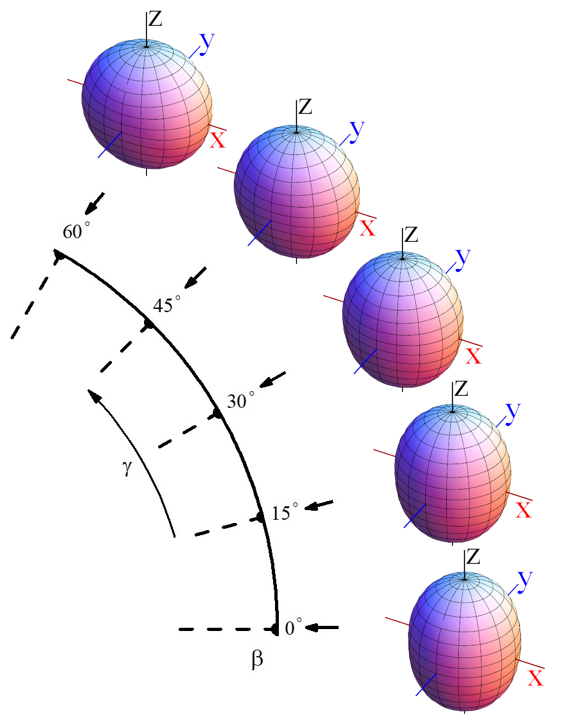}
		\label{fig:quadshapes2}}
	\caption{Nuclear shapes in the $(\beta,\ \gamma)$ plane. Panel (a) is from Ref.~\cite{Fortunato2005}.}
	\label{fig:quadshapes}
\end{figure}

Figure \ref{fig:quadshapes} displays nuclear shapes in the $(\beta,\ \gamma)$ plane. At $\gamma=0$ the nucleus is elongated along the $z$ axis, with the $x$ and $y$ semi-axes being equal. This axially symmetric shape is somewhat reminiscent of a cigar and is called {\em prolate}. As $\gamma$ increases, the $x$ semi-axis lengthens with respect the $y$ and $z$ semi-axes through a region of triaxial shapes with three unequal semi-axes, until axial symmetry is again reached at $\gamma=60^\circ$, but now with the $z$ and $x$ semi-axes equal in length. These two are longer than the $y$ semi-axis. The flat, pancake-like shape, is called {\em oblate}. This pattern is repeated: every $60^\circ$ axial symmetry recurs and prolate and oblate shapes alternate, but with the semi-axes permuted in their relative length.

\section{\label{sec:SO}Nuclear Surface Oscillations}
A simple framework that accurately describes the most prominent features of low-energy collective excitations is provided by the model of surface oscillations. Let the surface of the nucleus be parametrized by Eq.~(\ref{eq:surface}) and, if the coefficients $\alpha_{\lambda \mu}$ are small, the deformation potential energy and the associated kinetic energy take the form
\begin{equation}s
\label{eq:VVV}
V=\frac{1}{2} \sum_{\lambda \mu} C_{\lambda}\left|\alpha_{\lambda \mu}\right|^{2}, \quad \ \ T=\frac{1}{2} \sum_{\lambda \mu} B_{\lambda}\left|\dot{\alpha}_{\lambda \mu} \right|^{2},
\end{equation}
respectively. The dot denotes a derivative with respect to time, and the quantities $B_{\lambda}$ and $C_{\lambda}$ depend on the assumed properties of nucleonic  matter. For an incompressible nucleus of constant density $\varrho_0$, one finds $B_{\lambda}=\lambda^{-1} \varrho_{0} R_{0}^{5}$,
assuming nuclear matter to exhibit irrotational flow. If, moreover, the charge of the nucleus $Ze$ is uniformly distributed over its volume, one obtains
\begin{equation}
C_{\lambda}=(\lambda-1)(\lambda+2) R_{0}^{2} S-\frac{3}{2 \pi} \frac{\lambda-1}{2 \lambda+1} \frac{Z^{2} e^{2}}{R_{0}}
\end{equation}
where $S$ is the surface tension.

It is convenient to make a change of coordinates in such a way that the kinetic energy separates into a vibrational and a rotational part. This coordinate transformation will be considered in the following paragraphs for deformations of order $\lambda = 2$. For convenience, $B_2$ and $C_2$ are simplified to $B$ and $C$.

Consider a coordinate system $K^\prime$ whose axes coincide with the principal axes of the ellipsoid.
The coordinate transformation from the five $\alpha_\mu$ to the new coordinates $a_0$, $a_2$ and $\theta_i$ must be given by
\begin{equation}
\label{eq:alpha_mu}
\alpha_{\mu}=\sum_{\nu} a_{\nu} D_{\mu \nu}^{2*}\left(\theta_{i}\right) ,
\end{equation}
where $D_{\mu \nu}^{2*}\left(\theta_{i}\right)$ is the Wigner function and $\theta_i \equiv (\theta_1,\ \theta_2,\ \theta_3)$ are the three Euler angles which describe the orientation of the body-fixed axes in space.
Using Eq.~(\ref{eq:polarco}) and the unitary character of $D^2_{\mu \nu}$, one obtains
\begin{equation}
\label{eq:VVV_beta}
\sum_{\mu}\left|\alpha_{\mu}\right|^{2}=\sum_{\nu} a_{\nu}^{2}=a_{0}^{2}+2 a_{2}^{2}=\beta^{2} .
\end{equation}
From equations (\ref{eq:VVV}) and (\ref{eq:VVV_beta}), it follows $V=\frac{1}{2}C \beta^2$ for the potential energy of quadrupole deformation. To express the kinetic energy of the oscillating nucleus, from Eq.~(\ref{eq:alpha_mu}) one derives
\begin{equation}
\dot{\alpha}_{\mu}=\sum_{\nu} \dot{a}_{\nu} D_{\mu \nu}^{2*}\left(\theta_{i}\right)+\sum_{\nu j} a_{\nu} \dot{\theta}_{j} \frac{\partial}{\partial \theta_{j}} D_{\mu \nu}^{2*}\left(\theta_{i}\right).
   \end{equation}
If this expression is used in Eq.~(\ref{eq:VVV}), the kinetic energy can be written as a sum of three terms. The first term is quadratic in $\dot{a}_{\nu}$, and represents shape vibrations of the ellipsoid that retains its orientation. The second term, quadratic in $\dot{\theta}_{i}$, represents a rotation of the ellipsoid without change of its shape. The third term, which contains mixed time derivatives $\dot{a}_{\nu} \dot{\theta}_{i}$, vanishes as can be shown from simple properties of the Wigner coefficients $D^2_{\mu \nu}$ and their derivatives.

Therefore, the kinetic energy reads
\begin{equation}
\label{eq:Ttol}
T=T_{\mathrm{vib}}+T_{\mathrm{rot}}.
\end{equation}
For the vibrational energy, one obtains
\begin{equation}
\label{eq:Tvib}
T_{\mathrm{vib}}=\frac{1}{2} B \sum_{\nu}\left|\dot{a}_{\nu}\right|^{2}=\frac{1}{2} B\left(\dot{\beta}^{2}+\beta^{2} \dot{\gamma}^{2}\right)
\end{equation}
by means of Eq.~(\ref{eq:polarco}).
To derive a convenient form for $T_{\text{rot}}$, the most natural description is achieved if the principal axes of the nucleus determine the body-fixed system, because then the inertia tensor is diagonal. The classical rotational energy reads
\begin{equation}
\begin{aligned}
\label{eq:T_rot}
T_{\mathrm{rot}}&=\frac{1}{2} \sum_{i=1}^{3} \mathcal{J}_{i} \omega_{i}^{\prime  2},\\
\mathcal{J}_{i}&=4B\beta^2\sin^2(\gamma-i\frac{2\pi}{3}), \quad i=1,2,3
\end{aligned}
\end{equation}
where $\mathcal{J}_{i}$ is the moment of inertia about the $i$-th principal axis of the nucleus, and $\omega_{i}^\prime$ is the corresponding angular velocity of rotation.

The next step is the quantization of the classical Hamiltonian. As it is well known, there is no unique prescription for such a quantization in the general case. Usually one adopts the Pauli prescription, for which the Laplace operator is expressed in curvilinear coordinates. The final form of the Hamiltonian when expressed in $\beta$ and $\gamma$, as was done originally by Bohr \cite{Bohr1952}, reads
\begin{equation}\label{Hcoll}
\hat{H}=-\frac{\hbar^{2}}{2 B}\left[\frac{1}{\beta^{4}} \frac{\partial}{\partial \beta} \beta^{4} \frac{\partial}{\partial \beta}+\frac{1}{\beta^{2}} \frac{1}{\sin (3 \gamma)} \frac{\partial}{\partial \gamma} \sin (3 \gamma) \frac{\partial}{\partial \gamma}\right]+\sum_{i=1}^{3} \frac{\hat{I}_{i}^{2}}{2 \mathcal{J}_{i}}+\frac{1}{2}C \beta^2
\end{equation}
where $\hat{I_{i}}$ are operators of the projections of the nuclear angular momenta on the principal axes. However, the operators $\hat{I_{i}}$ cannot be identified with the standard angular-momentum operators in the laboratory frame. Their commutation relations look similar to those of the laboratory-fixed operators, but with a crucial change in sign:
\begin{equation}
\left[\hat{I}_{1}, \hat{I}_{2}\right]=-\mathrm{i} \hbar \hat{I}_{3}, \quad\left[\hat{I}_{2}, \hat{I}_{3} \right]=-\mathrm{i} \hbar \hat{I}_{1}, \quad\left[\hat{I}_{3}, \hat{I}_{1}\right]=-\mathrm{i} \hbar \hat{I}_{2}.
\end{equation}
Detailed expression for these operators can be found in Ref. \cite{Davydov1965}. 

The stationary wave functions separate in the following way:
\begin{equation}
\Psi\left(\beta, \gamma, \theta_i\right)=f(\beta) \Phi\left(\gamma, \theta_{i}\right).
\end{equation}
The solution proceeds through the familiar procedure of separation of variables and,
thus, the normalized solution for the $\beta$ part of the wave function reads
\begin{equation}
f_{n l}(\beta)= (-1)^{n} \left[\frac{2 (n !)^3}{\Gamma\left(n+l+\frac{5}{2}\right)}\right]^{1 / 2} \beta^{l} e^{-\beta^{2} / 2} L_{n}^{l+3 / 2}\left(\beta^{2}\right) ,
\end{equation}
where $n$ is the principal quantum number, while $l$ is the quantum number that corresponds to the solution of the eigenvalue equation of the Casimir operator of the $SO(5)$ group (also called $SO(5)$ seniority). The total energy $E$ of the five-dimensional quadrupole oscillator reads
\begin{equation}
E=(2n+l+\frac{5}{2})\hbar \omega=(N+\frac{5}{2})\hbar \omega, \quad N=0,1,\cdots
\end{equation}
The wave function $\Phi$ can be written in the form
\begin{equation}
\label{eq:wavegam}
\Phi_{M l \nu }^{I}\left(\gamma, \theta_{i}\right)=\sum_{K=-I}^{I} g_{K l \nu }^{I}(\gamma) D_{M K}^{I*}\left(\theta_{i}\right)
\end{equation}
where $I$, $M$, and $K$ are the quantum numbers for the total angular momentum, its projections on $Z$ axis in the laboratory frame, and on the third axis in the intrinsic frame, respectively. $\nu$ is a label that distinguishes between multiple sets of the same quantum numbers $I$, $M$ in a given $SO(5)$ irreducible representation. The functions $g$ are given explicitly in Ref.~\cite{Corrigan1976}.

In addition, the above expression must be symmetrized because of the ambiguity in the choice of intrinsic axes. The total number of possible choices is 24, and they can be related by three basic transformations $\hat{R}_1$, $\hat{R}_2$, $\hat{R}_3$, shown in Fig. \ref{fig:transformations}.
\begin{figure}[ht]
	\centering
	\subfigure[$\hat{R}_1$]{%
		\animategraphics[width=3.5cm,height=3.5cm, autoplay, loop]{1}{picture/fig3-transformations/R1-}{1}{2}
		\label{fig:R1}}
	\quad
	\subfigure[$\hat{R}_2$]{%
		\animategraphics[width=3.5cm,height=3.5cm, autoplay, loop]{1}{picture/fig3-transformations/R2-}{1}{4}
		\label{fig:R2}}
	\quad
	\subfigure[$\hat{R}_3$]{%
		\animategraphics[width=3.5cm,height=3.5cm, autoplay, loop]{0.5}{picture/fig3-transformations/R3-}{1}{3}
		\label{fig:R3}}
	\caption{Three basic transformations $\hat{R}_1$, $\hat{R}_2$ and $\hat{R}_3$ that define the symmetry of the collective wave function.}
	\label{fig:transformations}
\end{figure}
Under these transformation the collective wave functions must satisfy:
\begin{equation}
\begin{aligned}
&\hat{R}_{1}:\Psi\left(\beta, \gamma, \theta_{1}, \theta_{2}, \theta_{3}\right)=\Psi\left(\beta, \gamma, \theta_{1}+\pi, \pi-\theta_{2},-\theta_{3}\right),\\
&\hat{R}_{2}:\Psi\left(\beta, \gamma, \theta_{1}, \theta_{2}, \theta_{3}\right)=\Psi\left(\beta,-\gamma, \theta_{1}, \theta_{2}, \theta_{3}+\frac{\pi}{2}\right),\\
&\hat{R}_{3}:\Psi\left(\beta, \gamma, \theta_{1}, \theta_{2}, \theta_{3}\right)=\Psi\left(\beta, \gamma+\frac{2 \pi}{3}, \theta_{1}, \theta_{2}+\frac{\pi}{2}, \theta_{3}+\frac{\pi}{2}\right).
\end{aligned}
\end{equation}

By applying $\hat{R}_2^2$ to the wave function (\ref{eq:wavegam}), one obtains that $K$ must be even.
$\hat{R}_1$ will transform  $D_{ M K}^{I*}(\mathbf{\theta})$ into $(-1)^I D_{ M -K}^{I*}(\mathbf{\theta})$, and require that $g_{-K l \nu}^I (\gamma)=(-1)^Ig_{K l \nu  }^I (\gamma)$. Finally, the wave function is written to explicitly display the symmetry property which restricts the sum to $K \ge 0$:
\begin{equation}
\label{eq:bohrwave}
\begin{aligned}
\Psi_{M n l \nu  }^I \left(\beta, \gamma, \theta_{i}\right) &={N}_{n l \nu  }^I f_{n l}(\beta) \sum_{K \geq 0 \atop \text { even }} g_{K l \nu  }^I (\gamma)\phi_{M K}^I\left(\theta_{i}\right),
\end{aligned}
\end{equation}
where
\begin{equation}
\label{eq:phi}
\phi_{M K}^I\left(\theta_{i}\right)=\sqrt{\frac{2 I+1}{16(1+ \delta_{K 0}) \pi^2}}\left[D_{M K}^{I *}(\theta_{i})+(-1)^{I} D_{M,-K}^{I *}(\theta_{i})\right]
\end{equation}
are the properly symmetrized wave functions for the symmetric rotor. ${N}_{n l \nu  }^I$ normalizes the $\gamma$ part of Eq.~(\ref{eq:bohrwave}). Detailed expression for the wave function can be found in Ref.~\cite{Corrigan1976}.

\section{\label{sec:RVM}The Rotation-Vibration Model}

\begin{figure}
	\centering
	\includegraphics[width=1\linewidth]{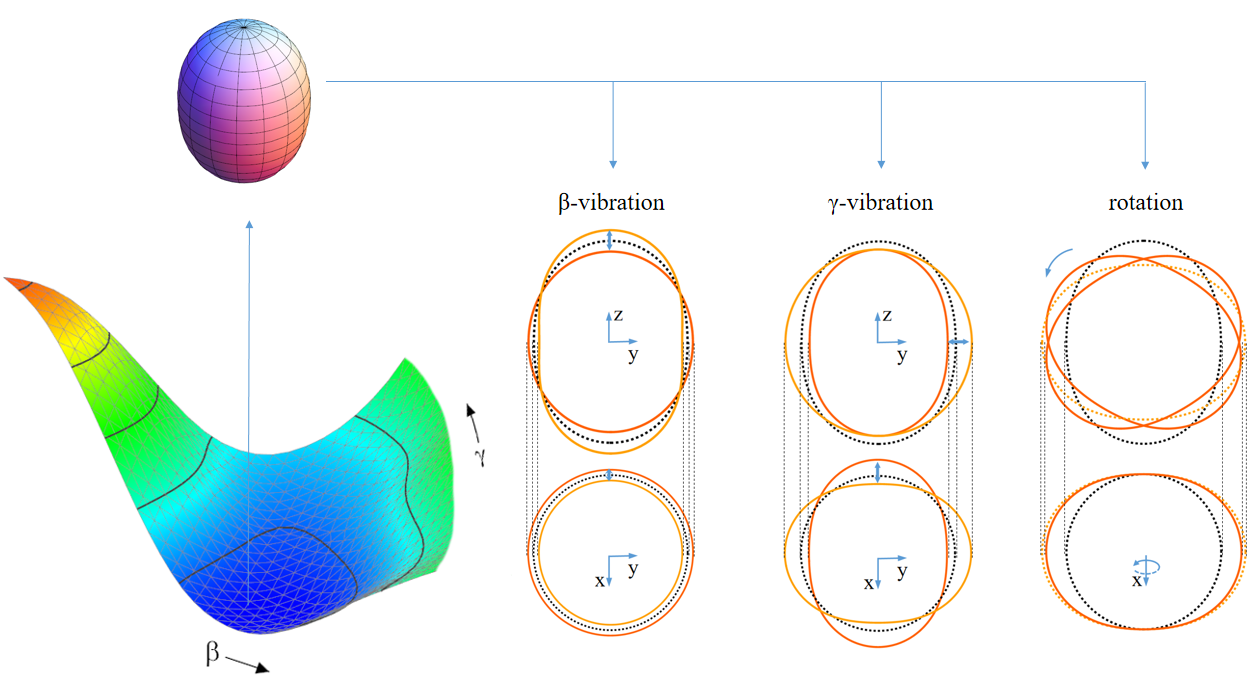}
	\caption{Schematic plot of the collective potential $V(\beta,\ \gamma)$ for $0 \leq \gamma \leq \pi/3$, shape of global minimum, vibrations in $\beta$ and $\gamma$ directions, and rotation.}
	\label{fig:potential}
\end{figure}

 The most important special case of collective surface motion is that of well deformed nuclei, whose potential energy surfaces exhibit a deep axially deformed minimum like in the simple rotor model, but with the additional feature of small oscillations around that minimum in both $\beta$ and $\gamma$ degrees of freedom (see the illustration in Fig. \ref{fig:potential}).
As the harmonic vibrations are easier to deal with in Cartesian coordinates, the potential is expressed in the coordinates $(a_0,a_2)$ instead of $(\beta, \gamma)$. Assuming the potential minimum is located at $a_0 = \beta_0$ and $a_2 = 0$, the small displacement from the equilibrium deformation can be written in the form
\begin{equation}
\xi=a_{0}-\beta_{0}, \quad \eta=a_{2}=a_{-2}
\end{equation}
and it is easy to show that the Hamiltonian can be expressed in the form
\begin{equation}
H=T_{\text{vib}}+T_{\text{rot}}+V=\frac{1}{2} B\left(\dot{\xi}^{2}+2 \dot{\eta}^{2}\right)+\frac{1}{2} \sum_{i=1}^{3} \mathcal{J}_{i} \omega_{i}^{\prime 2}+\frac{1}{2} C_{0} \xi^{2}+C_{2} \eta^{2}.
\end{equation}
Note the additional factor of $2$ in the $\eta$-dependent part of the potential, which takes into account that $\eta$ represents the two coordinates $a_2$ and $a_{-2}$.

In a quantum mechanical description, a nucleus cannot rotate around the symmetry axis. However, here the vibrations make such rotations possible dynamically, and they are coupled to the dynamic fluctuations from axial symmetry described by $\eta$. The moments of inertia are given by the lowest order expressions
\begin{equation}
\mathcal{J} \equiv \mathcal{J}_{1}=\mathcal{J}_{2}=3 B \beta_{0}^{2}, \quad \mathcal{J}_{3}=8 B \eta^{2},
\end{equation}
and the simplest expression for the kinetic energy reads
\begin{equation}
T=\frac{1}{2} B\left(\dot{\xi}^{2}+2 \dot{\eta}^{2}\right)+\frac{1}{2} \mathcal{J}\left(\omega_{1}^{\prime 2}+\omega_{2}^{\prime 2}\right)+4 B \eta^{2} \omega_{3}^{\prime 2} .
\end{equation}
Similarly, the quantized Hamiltonian can be written in the form
\begin{equation}
\label{eq:Hcoll}
\hat{H}(\xi, \eta, \boldsymbol{\theta})= \frac{-\hbar^{2}}{2 B}\left(\frac{\partial^{2}}{\partial \xi^{2}}+\frac{1}{2} \frac{\partial^{2}}{\partial \eta^{2}}\right)
+\frac{\hat{I}_{1}^{ 2}+\hat{I}_{2}^{ 2}}{2 \mathcal{J}}+\frac{\hat{I}_{3}^{ 2}-\hbar^{2}}{16 B \eta^{2}}+\frac{1}{2} C_{0} \xi^{2}+C_{2} \eta^{2} .
\end{equation}
The term with $\eta^{-2}$ is similar to the centrifugal potential and can be treated accordingly. The eigenfunctions of the rotational operator in the Hamiltonian correspond to those of the rigid rotor, while the $\xi$-dependent and $\eta$-dependent parts can obviously be separated. Thus, a trial wave function takes the form
\begin{equation}
\psi(\xi, \eta, \boldsymbol{\theta})= g(\xi) \chi(\eta) {D}_{M K}^{I *}(\theta_i) .
\end{equation}
The next step is to complete the separation of the $\xi$ and $\eta$ dependence, and finally they correspond to typical eigenvalue problems with one-dimensional and three-dimensional harmonic oscillator potentials, respectively. The total energy reads
\begin{equation}
E_{n_{\beta} n_{\gamma} I K}=\hbar \omega_{\beta}\left(n_{\beta}+\frac{1}{2}\right)+\hbar \omega_{\gamma}\left(2 n_{\gamma}+\frac{1}{2}|K|+1\right)+\frac{\hbar^{2}}{2 \mathcal{J}}\left[I(I+1)-K^{2}\right] ,
\end{equation}
with
\begin{equation}
	\omega_{\beta}=\sqrt{\frac{C_0}{B}}, \quad \omega_{\gamma}=\sqrt{\frac{C_2}{B}}, \quad n_{\beta} \ \text{or} \  n_{\gamma}=0,1,\cdots
\end{equation}
For the eigenfunctions that depend on the $\xi$ ($\beta$) coordinate, the one-dimensional harmonic oscillator eigenfunctions can be used directly, and they are simply written as $\langle \xi|n_\beta \rangle$. For the $\gamma$ coordinate the solution for the three-dimensional harmonic oscillator potential is used
\begin{equation}
\chi_{K n_{\gamma}}(\eta)=N_{K n_{\gamma}} \sqrt{|\eta|} \eta^{K / 2} \mathrm{e}^{-\lambda \eta^{2} / 2}{ }_{1} F_{1}\left(-n_{\gamma}, l_{K}+\frac{3}{2}, \lambda \eta^{2}\right) ,
\end{equation}
with $\lambda=2 B \omega_{\gamma} / \hbar$, $\l_K=(-1\pm K)/2$,
and $N_{K n_{\gamma}}$ is the normalization factor.
After symmetrization, the wave functions read
\begin{equation}
\psi^I_{ M K n_{\beta} n_{\gamma}}(\xi, \eta, \boldsymbol{\theta})=\left\langle\xi \mid n_{\beta}\right\rangle \chi_{K n_{\gamma}}(\eta)\phi_{M K}^I\left(\theta_{i}\right) ,
\end{equation}
and $\phi_{M K}^I\left(\theta_{i}\right)$ are the symmetrized eigenfunctions for the symmetric rotor of Eq. (\ref{eq:phi}).
The allowed values of the quantum numbers are:
\begin{equation}
\begin{aligned}
K &=0,2,4, \ldots \\
I &= \begin{cases}K, K+1, K+2, \ldots \ \ \ \ \text { for } K \neq 0 \\
0,2,4, \ldots \ \ \ \ \ \ \ \ \ \ \ \ \ \ \ \ \ \ \ \text { for } K=0\end{cases} \\
M &=-I,-I+1, \ldots,+I.
\end{aligned}
\end{equation}

The structure of a typical spectrum is shown in Fig.~\ref{Spectrum-habook}. The bands are characterized by a given set of $K,n_\beta,n_\gamma$ and the excitation energies follow the $I(I+1)$ rule of a rotor. The principal bands are:
\begin{figure*}[ht]
\centering{\includegraphics[width=1\linewidth]{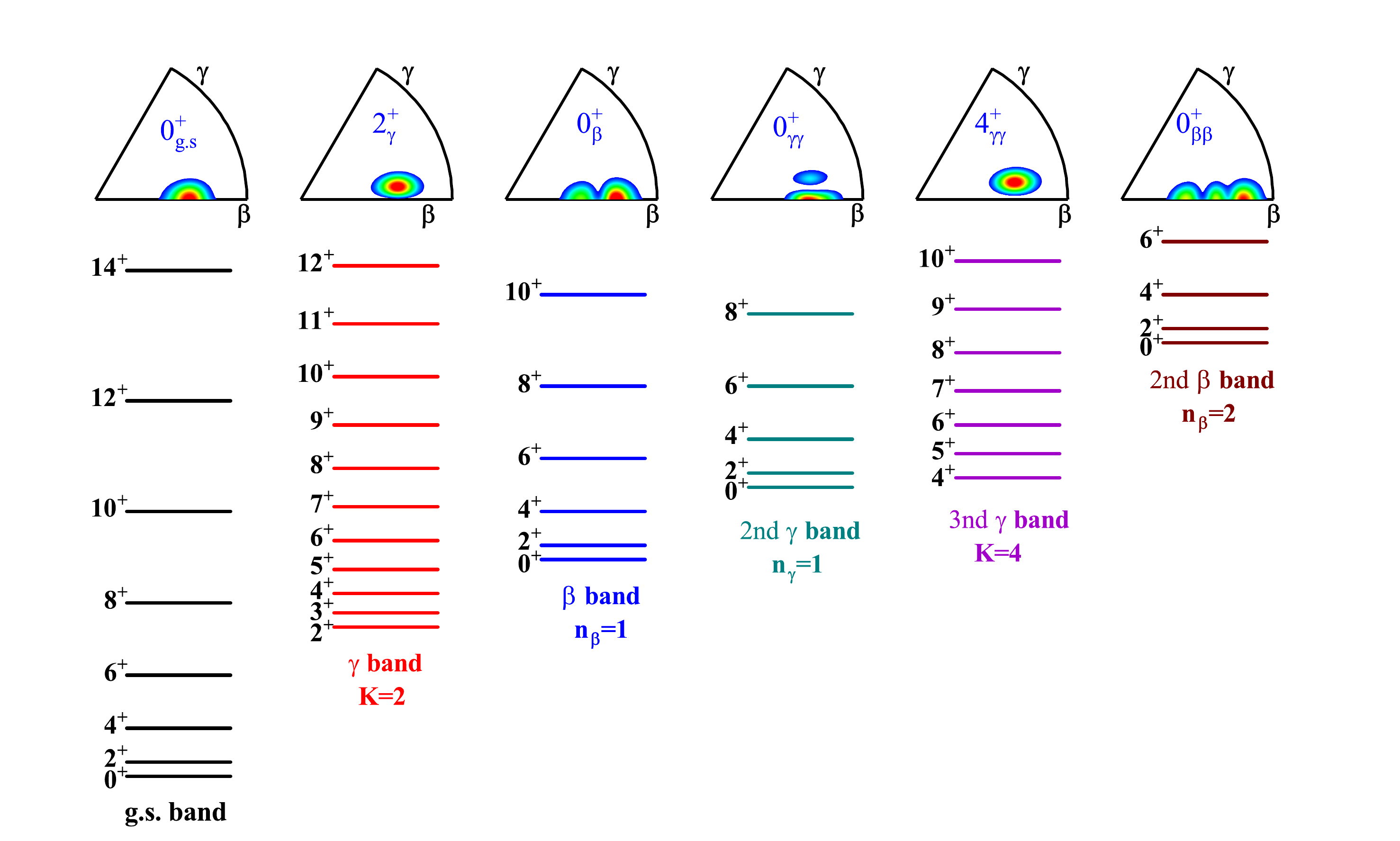}}
\caption{\label{Spectrum-habook} Structure of the excitation spectrum of the rotation-vibration model. The probability density distributions in the $(\beta,\ \gamma)$ plane for the band-heads are shown above each band. The notation and the corresponding quantum numbers are indicated below the band-heads.}
\end{figure*}
\begin{itemize}
    \item the ground-state band (g. s.), includes the states $|IM000\rangle$ with $I$ even. The excitation energies are given by $\hbar^ 2 I(I+1)/2\mathcal J$.
    \item the $\beta$ band, includes the states $|IM010\rangle$ with one quantum of vibration in the $\beta$ direction. It starts at $\hbar\omega_\beta$ above the ground state, and also contains only even angular momenta.
    \item the $\gamma$ band corresponds to $K=2$. It is characterized by coupling between rotation (around the third principal axis) and $\gamma$-vibration, induced by the term $(\hat{I}_{3}^{ 2}-\hbar^{2})/16 B \eta^{2}$ in the Hamiltonian. In the spectra it is easy to distinguish it from the $\beta$ band, since it starts with $2^+$ and contains both even and odd angular momenta.
    \item the next higher bands are the $\gamma$ band with $n_\gamma = 1$ and the one with $ K = 4$, the second $\beta$ band with $n_\beta=2$.
\end{itemize}

The simplest observables that can be calculated and compared to data, are the electric quadrupole moments and transition probabilities. The collective quadrupole operator in the intrinsic system (in lowest order of deformation parameters) is given by
\begin{equation}
\hat{Q}_{2 \mu}= \frac{3 Z e}{4 \pi} R_{0}^{2}\left\{{D}_{\mu 0}^{2 *}(\theta_i)(\beta_{0}+\xi)
+\left({D}_{\mu 2}^{2 *}(\theta_i)+{D}_{\mu-2}^{2 *}(\theta_i)\right) \eta\right\}.
\end{equation}
The collective quadrupole moment is defined by
\begin{equation}
Q_{I K n_{\beta} n_{\gamma}}=\sqrt{\frac{16 \pi}{5}}\left\langle I M=I K n_{\beta} n_{\gamma}\left|\hat{Q}_{20}\right| I M=I K n_{\beta} n_{\gamma}\right\rangle .
\end{equation}
It does not depend on $n_\beta$ and $n_\gamma$, and its value reads
\begin{equation}
Q_{I K}=Q_{0} \frac{3 K^{2}-I(I+1)}{(I+1)(2 I+3)} ,
\end{equation}
where the intrinsic quadrupole moment is defined
\begin{equation}
Q_{0}=\frac{3 Z e R_{0}^{2}}{\sqrt{5 \pi}} \beta_{0}.
\end{equation}
The reduced electric quadrupole transition probabilities are calculated using the expression
\begin{equation}
B\left(E 2 ; I_{i} \rightarrow I_{f}\right)=\frac{1}{2 I_{i}+1}\left|\left\langle I_{f}|| \hat{Q}|| I_{i} \right\rangle\right|^{2} .
\end{equation}
The so-called stretched $BE2$-values in a rotational band read 
\begin{equation}
B(E 2; I+2 \rightarrow I)=Q_{0}^{2} \frac{5}{16 \pi}\left|C_{K\ \ \ \ 0\ K}^{I+2\ 2\ I} \right|^{2}
\end{equation}
and, for $K = 0$ bands, one obtains
\begin{equation}
B(E 2, I+2 \rightarrow I)=Q_{0}^{2} \frac{5}{16 \pi} \frac{3}{2} \frac{(I+1)(I+2)}{(2 I+3)(2 I+5)}.
\end{equation}
The transition probabilities allow the determination of the intrinsic quadrupole moment and thus the deformation itself, whereas the energy spacings in the excitation spectra determine the moment of inertia.

%
\section{\label{sec:GCM+GOA}   Microscopic Derivation of the Collective Hamiltonian}
%

The general form of the collective Bohr Hamiltonian has been derived starting from a microscopic Hamiltonian or from an energy density functional in two rather different ways: (i) using the generator coordinate method (GCM) \cite{Haff1973, Banerjee1973, Giraud1974, Onishi1975, Ring1980}, and (ii) applying the time-dependent Hartree-Fock (TDHF) theory \cite{Baranger1968,Baranger1978,Goeke1978}. Here, the former one is presented, taking into account  that the derivation based on the GCM is fully quantum-mechanical. It relies on the validity of the Gaussian overlap approximation (GOA) for the overlaps between configurations with different deformations, and on the assumption that the collective velocities are small, i.e., that the expansion in the collective momenta up to second order is adequate. The validity of these approximations is  demonstrated in a comparison with a full GCM calculation for a nucleus characterized by shape coexistence: $^{76}$Kr \cite{Yao2014}.

\subsection{\label{GCM+GOA:sec1} General Concepts of the Generator Coordinate Method}

\begin{figure}[htp]
\begin{center}
\includegraphics[width=1.0\textwidth]{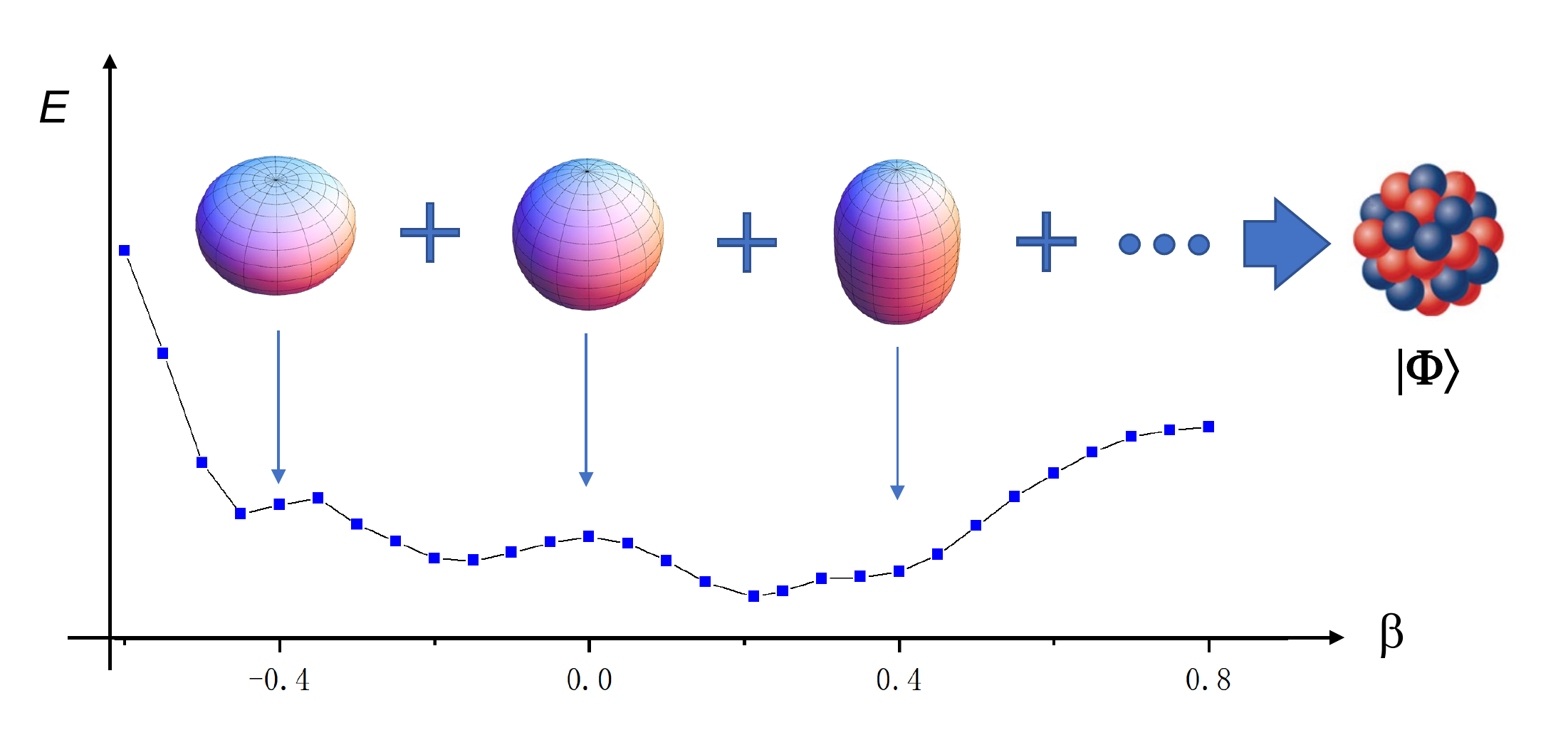}
\caption{\label{fig:GCM}
The nuclear wave function $|\Phi\rangle $ as a continuous superposition of generating functions $|\varphi(\beta )\rangle$, labeled by the parameters $\beta$.}
\end{center}
\end{figure}

In the generator coordinate method, the nuclear wave function $|\Phi\rangle $ can be expressed as a continuous superposition of generating functions $|\varphi(\beta )\rangle$, that are labeled by an arbitrary number of real or complex parameters $\{\beta\}=\beta_1, \beta_2, \ldots, \beta_i$, the so-called generator coordinates,
\begin{equation}
    |\Phi\rangle =\int d\beta f(\beta )|\varphi(\beta )\rangle.  \label{GCMexpand}
\end{equation}
The weight $f(\beta)$ is assumed to be a well behaved function of the variables $\beta$, and the multidimensional integral includes all real and imaginary parameters $\beta$. The set of generating wave functions $\varphi(\beta)$ is often determined by a specific problem or geometry. In many cases,  the intrinsic states of a nucleus are chosen as the generating states, and the corresponding shape deformation parameters as the generator coordinates. See the schematic illustration in figure \ref{fig:GCM}.

The variational principle implies
\begin{equation}
    \begin{split}
        \delta E(\Phi)=\delta \frac{\langle \Phi|\hat H|\Phi\rangle }{\langle \Phi|\Phi\rangle }
                      =\frac{1}{\langle \Phi|\Phi\rangle }\left(\langle \Phi|\hat H-E|\delta\Phi\rangle +\langle \delta\Phi|\hat  H-E|\Phi\rangle \right)=0.      \label{variation principle}
    \end{split}
\end{equation}
By inserting Eq.(\ref{GCMexpand}) into Eq.(\ref{variation principle}), and performing the variation with respect to $f(\beta)$, an integral equation is obtained
\begin{gather}
    \int \langle \varphi(\beta)|\hat H|\varphi(\beta ^\prime)\rangle f(\beta ^\prime)d\beta^\prime=E\int \langle \varphi(\beta)|\varphi(\beta^\prime)\rangle f(\beta^\prime)d\beta^\prime. \label{H-W equ}
\end{gather}
This is the \emph{Hill-Wheeler equation} and it can formally be written in the form
\begin{equation}
     Hf=E N f,
\end{equation}
with the overlap functions
\begin{gather}
    \begin{split}
        H (\beta ,\beta ^\prime)&=\langle \varphi(\beta )|\hat H|\varphi(\beta ^\prime)\rangle~~~~~~ The~energy~kernel\\
        N (\beta ,\beta ^\prime)&=\langle \varphi(\beta )|\varphi(\beta ^\prime)\rangle~~~~~~~~~~ The~overlap~kernel
    \end{split}
\end{gather}
as integral kernels.

To solve this equation, a transformation is defined from the space of the generating coordinates
to the coordinate space in which the wave functions specified by the true coordinates $\eta$ are orthogonal. The transformation $\beta\rightarrow \eta$ is defined in such a way that $N(\beta,\beta^\prime)\rightarrow\delta(\eta-\eta^\prime)$.

In the first step, the eigenvalue equation for the overlap kernel is solved
\begin{gather}
    \int N(\beta,\beta^\prime)\chi (s,\beta^\prime)d\beta^\prime=\nu(s)\chi(s,\beta) .
    \label{eigenN}
\end{gather}
The eigenfunctions specified by a new variable $s$ form a complete orthonormalized set
\begin{gather}
    \int\chi^*(s,\beta)\chi(s^\prime,\beta)d\beta=\delta(s-s^\prime); \ \ \ \ \
    \int\chi^*(s,\beta)\chi(s,\beta^\prime)ds=\delta(\beta-\beta^\prime) .
\end{gather}
Next, a transformation kernel is introduced
\begin{gather}
    N^{\frac{1}{2}}(\beta,\eta)=\int\chi(s,\beta)\nu^{\frac{1}{2}}(s)\chi^*(s,\eta)ds  ,
    \label{N1/2 def}
\end{gather}
that represents the transition from $\delta(\eta-\eta^\prime)$ to $N(\beta,\beta^\prime)$ through
\begin{eqnarray}
        \int N^{\frac{1}{2}*}(\beta,\eta)\delta(\eta-\eta^\prime)N^{\frac{1}{2}}(\eta^\prime,\beta^\prime)d\eta d\eta^\prime
        &=&\int \chi^*(s,\beta^\prime)\nu(s)\chi(s,\beta)ds\nonumber\\
        &=&\int N(\beta,\beta^{\prime\prime})\delta(\beta^\prime-\beta^{\prime\prime})d\beta^{\prime\prime}\nonumber\\
        &=&N(\beta,\beta^\prime)  . 
        \label{delta to N}
\end{eqnarray}
Obviously, the inverse $N^{-\frac{1}{2}}$ has to be determined to transform $N(\beta,\beta^\prime)$ to $\delta(\eta-\eta^\prime)$.
Accordingly, the energy kernel should be transformed to the coordinate representation by using the narrowing kernel
\begin{gather}
    L(\eta,\eta^\prime)=\int N^{-\frac{1}{2}*}(\beta,\eta)H(\beta,\beta^\prime)N^{-\frac{1}{2}}(\beta^\prime,\eta^\prime)d\beta d\beta^\prime  ,
\label{narrowing energy kernel def}
\end{gather}
and the new wave function $g(\eta)$ in coordinate representation is related to the generating function $f(\beta)$
\begin{gather}
    g(\eta)=\int N^{\frac{1}{2}}(\beta,\eta)f(\beta)d\beta . 
\label{g def}
\end{gather}
By replacing each factor in Eq.(\ref{H-W equ}) with the corresponding one after transformation, the orthogonal Hill-Wheeler equation is obtained
\begin{gather}
    \int L(\eta,\eta^\prime)g(\eta^\prime)d\eta^\prime=Eg(\eta) ,
    \label{ND New HW}
\end{gather}
or in matrix form
\begin{gather}
    Lg=Eg .
\end{gather}
Eq. (\ref{ND New HW}) can be considered as a stationary Schr\"{o}dinger equation, where $L$ is the collective Hamiltonian and $g$ the corresponding eigenfunction in the new representation. To derive the collective Hamiltonian, one needs an explicit expression for the transformation kernel $N^\frac{1}{2}$. This is achieved by introducing the Gaussian overlap approximation.

\subsection{\label{GCM+GOA:sec2} The Gaussian Overlap Approximation}

A simple and pedagogical derivation of the collective Hamiltonian can be presented by using a single generator coordinate. A set of time-reversal invariant generating functions $|\beta\rangle =|\varphi(\beta)\rangle $, which depend on a single collective parameter $\beta$, can be obtained by constrained self-consistent mean-field calculations. To derive the collective Hamiltonian, the Gaussian overlap approximation (GOA) is based on two assumptions: 1) generally, the overlap function $N =\langle \beta|\beta^\prime\rangle $ rapidly becomes smaller with the increase of the distance $|\beta-\beta^\prime|$; and 2) both the energy kernel $H$ and overlap kernel $N$ are well behaved functions. Therefore, 
a Gaussian function is assumed for the overlap kernel $N$ 
\begin{gather}
     N (\beta ,\beta ^\prime)= e^{-\frac{1}{2}\gamma_0(\beta -\beta ^\prime)^2} .
     \label{GOA}
\end{gather}
Note that Eq.(\ref{GOA}) is written in a homogeneous form, namely $\gamma_0$ is constant.

The overlap kernel is a function that only depends on the difference $(\beta -\beta ^\prime)$. In this case, the solutions of equation (\ref{eigenN}) are
\begin{gather}
    \nu(s)=\sqrt{\frac{2\pi}{\gamma_0}}e^{-\frac{s^2}{2\gamma_0}}; \ \ \ \ \
    \chi(s,\beta )=\frac{1}{\sqrt{2\pi}}e^{is\beta} ,
\end{gather}
for the eigenvalue and eigenfunction, respectively. Hence, the kernel $N^{\frac{1}{2}}$ can be expressed as
\begin{gather}
        N^{\frac{1}{2}}(\beta,\eta)=\int \chi^*(s,\beta) \nu^\frac{1}{2}(s)\chi(s,\eta)ds = (\frac{2\gamma_0}{\pi})^{\frac{1}{4}}e^{-\gamma_0 (\beta -\eta)^2} .
        \label{Nhalf def}
\end{gather}
Starting from Eq. (\ref{narrowing energy kernel def}), and using the expressions above for $\chi(s,\beta)$, $\nu(s)$, and $N^{-\frac{1}{2}}$, N. Onishi and  T. Une derived the collective Hamiltonian in  \cite{Onishi1975}. Here, the approach proposed by P. Ring and P. Schuck in their textbook  {\em The Nuclear Many-Body Problem} \cite{Ring1980} will be described.

One starts from the expectation of the Hamiltonian
\begin{gather}
    \label{exp E}
    E=\langle \Phi|\hat{H}|\Phi\rangle
    =\int f^*(\beta) H(\beta, \beta^\prime) f(\beta^\prime)d\beta d\beta^\prime 
\end{gather}
and, to transform $f(\beta)$ and $H(\beta, \beta^\prime)$ to $g(\eta)$ and $L(\eta, \eta^\prime)$ as in Eq. (\ref{ND New HW}), the reduced energy kernel $h(\beta, \beta^\prime)=H(\beta, \beta^\prime)/N(\beta, \beta^\prime)$ is introduced. The diagonal element of the reduced energy kernel is just the expectation value of the Hamiltonian in the state $|\beta\rangle$. Then Eq. (\ref{exp E}) takes the form
\begin{gather}
    \label{exp E1}
    E=\int f^*(\beta) N(\beta, \beta^\prime)h(\beta, \beta^\prime) f(\beta^\prime)d\beta  d\beta ^\prime .
\end{gather}
Inserting Eq. (\ref{delta to N}) into Eq. (\ref{exp E1}), the following expression is obtained
\begin{gather}
    \label{exp E2}
    E=\int f^*(\beta) N^{\frac{1}{2}*}(\beta,\eta) \delta(\eta-\eta^\prime)  h(\beta, \beta^\prime)N^{\frac{1}{2}}(\eta^\prime,\beta^\prime) f(\beta^\prime)d\beta  d\beta ^\prime d\eta d\eta^\prime .
\end{gather}
Next, $h(\beta, \beta^\prime)$ is expanded to second order around the point $\beta=\eta$, $\beta^\prime=\eta^\prime$ 
\begin{gather}
  \begin{split}
     h(\beta, \beta^\prime)&=h+h_\beta(\beta-\eta)+h_{\beta^\prime}(\beta^\prime-\eta^\prime)\\
                           &\ \ \ +\frac{1}{2}\left[h_{\beta \beta}(\beta-\eta)^2+2h_{\beta\beta^\prime}(\beta-\eta )(\beta^\prime-\eta^\prime)+h_{\beta^\prime\beta ^\prime}(\beta^\prime-\eta^\prime )^2\right] .\label{expand h}
  \end{split}
\end{gather}
To calculate the derivatives $h$, the Hamiltonian kernel and overlap kernel are written in the form
\begin{gather}
  \begin{split}
    H(\beta,\beta^\prime)&=\langle\eta|e^{-i(\beta-\eta)\hat{P}/\hbar}\hat{H}e^{i(\beta^\prime-\eta^\prime)\hat{P}/\hbar}|\eta^\prime\rangle\\
    N(\beta,\beta^\prime)&=\langle\eta|e^{-i(\beta-\eta)\hat{P}/\hbar}e^{i(\beta^\prime-\eta^\prime)\hat{P}/\hbar}|\eta^\prime\rangle
  \end{split}
\end{gather}
where $\hat{P}=-i\hbar\frac{\partial}{\partial\beta}$ is the momentum operator. Then the derivative $h$ can be expressed as
\begin{gather}
    h_{\beta^\prime}=h_{\beta}=\left. \frac{\partial}{\partial\beta}\left(\frac{H(\beta,\beta^\prime)}{N(\beta,\beta^\prime)}\right)\right|_{\beta=\beta^\prime=\eta=\eta^\prime}=\left.-\frac{i}{\hbar}\langle\eta|\hat{P}\hat{H}|\eta^\prime\rangle\right|_{\eta=\eta^\prime}=0 ,
\end{gather}
where $\eta=\eta^\prime$ because of $\delta(\eta-\eta^\prime)$ in Eq. (\ref{exp E2}). Note that the generating functions $|\eta\rangle$ are time-reversal invariant states, and $\hat{P}$ is a time-odd operator. Therefore, the first derivatives vanish, while the second derivatives read 
\begin{gather}
    \begin{split}
        h_{\beta^\prime\beta^\prime}&=h_{\beta\beta}=\left.-\frac{1}{\hbar^2}\left(\langle\eta|\hat{P}^2\hat{H}|\eta^\prime\rangle-\langle\eta|\hat{H}|\eta^\prime\rangle\langle\eta|\hat{P}^2|\eta^\prime\rangle \right)\right|_{\eta=\eta^\prime}\\
        h_{\beta\beta^\prime}&=\left.\frac{1}{\hbar^2}\left(\langle\eta|\hat{P}\hat{H}\hat{P}|\eta^\prime\rangle-\langle\eta|\hat{H}|\eta^\prime\rangle\langle\eta|\hat{P}^2|\eta^\prime\rangle \right)\right|_{\eta=\eta^\prime}  .
    \end{split}
\end{gather}
In addition, the term $\frac{\partial }{\partial\eta}h_{\beta}=h_{\beta\beta}+h_{\beta\beta^\prime}$ will be used when performing partial integration.
Setting $N_1=N^{\frac{1}{2}}(\beta ,\eta)$, $N_2=N^{\frac{1}{2}}(\beta^\prime,\eta^\prime)$, and inserting the expansion of $h$ into Eq. (\ref{exp E2}), one obtains
\begin{gather}
    \begin{split}
        E &=\int d\beta d\beta^\prime d\eta d\eta^\prime f^*(\beta )f(\beta^\prime)N_1N_2
            \Big\{h+h_\beta(\beta-\eta)+h_{\beta^\prime}(\beta^\prime-\eta^\prime)\\
        &\ \ \ \ \ +\frac{1}{2}\big[h_{\beta \beta}(\beta-\eta)^2+2h_{\beta\beta^\prime}(\beta-\eta ) (\beta^\prime-\eta^\prime)+h_{\beta^\prime\beta ^\prime}(\beta^\prime-\eta^\prime )^2\big]\Big\} .
        \label{Exp E3}
    \end{split}
\end{gather}
From Eq.(\ref{Nhalf def}), $(\beta -\eta)N_1$ and $(\beta -\eta)^2N_1$ can be expressed in the form
\begin{gather}
    \label{derN1}
    \begin{split}
    (\beta -\eta ) N_1&=\frac{1}{2\gamma_0}\frac{\partial}{\partial \eta } N_1\\
    (\beta -\eta )^2 N_1&=(\frac{1}{4\gamma_0^2}\frac{\partial^2}{\partial \eta ^2}+\frac{1}{2\gamma_0}) N_1 .
    \end{split}
\end{gather}
Obviously, $(\beta ^\prime-\eta^\prime)N_2$ and $(\beta ^\prime-\eta^\prime)^2N_2$ have the same forms as in Eq. (\ref{derN1}), except that the derivatives are with respect to $\eta^\prime$. Then Eq. (\ref{Exp E3}) becomes 
\begin{equation}
	\begin{aligned}
	\label{e1}
	E = &\int d\beta d\beta^\prime d\eta d\eta^\prime \delta(\eta-\eta^\prime)f^{*}(\beta )f(\beta^\prime)
	\left[(h+\frac{h_{\beta\beta}+h_{\beta^\prime\beta^\prime}}{4\gamma_0}) N_1 N_2+\frac{1}{2\gamma_0}h_\beta  N_1^\prime N_2\right.\\
	&\left.+\frac{1}{2\gamma_0}h_{\beta^\prime} N_1 N_2^\prime+\frac{1}{8\gamma_0^2}h_{\beta \beta } N_1^{\prime\prime} N_2+\frac{1}{4\gamma_0^2}h_{\beta \beta ^{\prime}} N_1^\prime N_2^\prime+\frac{1}{8\gamma_0^2}h_{\beta^\prime \beta^\prime } N_1 N_2^{\prime\prime}\right] .
	\end{aligned}
\end{equation}
After partial integration, and inserting Eq. (\ref{g def}), the final result is obtained neglecting higher-order derivatives 
\begin{gather}
        E=\int g^*\delta(\eta-\eta^\prime)\left[h-\frac{h_{\beta\beta ^\prime}}{2\gamma_0}+\frac{\partial}{\partial\eta }\frac{1}{4\gamma_0^2}(h_{\beta \beta }-h_{\beta \beta ^\prime})\frac{\partial}{\partial \eta }\right]gd\eta d\eta^\prime .   
        \label{finalltHvalue}
\end{gather}
Comparing to Eq. (\ref{ND New HW}), the collective Hamiltonian $L(\eta, \eta^\prime)$  finally reads
\begin{gather}
    \begin{split}
        L(\eta,\eta^\prime)&=\delta(\eta-\eta^\prime)\left[h-\frac{h_{\beta\beta ^\prime}}{2\gamma_0}+\frac{\partial}{\partial\eta }\frac{1}{4\gamma_0^2}(h_{\beta \beta }-h_{\beta \beta ^\prime})\frac{\partial}{\partial \eta }\right]\\
        &=\delta(\eta-\eta^\prime)\left[-\frac{\partial}{\partial \eta}\frac{1}{2M(\eta)}\frac{\partial}{\partial \eta}+V(\eta)\right],
        \end{split}
\end{gather}
with the collective mass and potential
\begin{gather}
    \label{collMV}
    \begin{split}
        \frac{1}{M(\eta)}&=\frac{-1}{2\gamma^2_0}(h_{\beta \beta }-h_{\beta \beta ^\prime}) , \\
        V(\eta)&=h-\frac{h_{\beta \beta ^\prime}}{2\gamma_0} ,
    \end{split}
\end{gather}
respectively.
In a more general case, namely for inhomogeneous GOA, $\gamma_0$ in Eq. (\ref{GOA}) depends on the collective coordinates. However, the inhomogeneous GOA can always be brought into a  homogeneous form by performing a transformation from $\beta$ to a new coordinate
$\alpha=\int^\beta \sqrt{\gamma(\beta^\prime)/\gamma_0}d\beta^\prime$ \cite{Ring1980}. Finally, to obtain the general form of the collective Hamiltonian, a transformation back to the original coordinate is carried out by inserting $d\eta= \sqrt{\gamma/\gamma_0}d\xi$ into Eq. (\ref{finalltHvalue}), and using a new wave function
\begin{gather}
\tilde{g}=\gamma_0^{-1/4}g=\int (\frac{2}{\pi})^{\frac{1}{4}}e^{-\gamma (\beta -\xi)^2}f(\beta)d\beta .
\end{gather}
Then, 
\begin{gather}
        E=\int \sqrt{\gamma}\tilde{g}^*\delta(\xi-\xi^\prime)\left[h-\frac{h_{\beta\beta ^\prime}}{2\gamma}+\frac{1}{\sqrt{\gamma}}\frac{\partial}{\partial\xi }\sqrt{\gamma}\frac{1}{4\gamma^2}(h_{\beta \beta }-h_{\beta \beta^\prime})\frac{\partial}{\partial \xi }\right]\tilde{g}d\xi d\xi^\prime    \label{finalltHvalue2},
\end{gather}
Note that $h_{\beta \beta }$ and $h_{\beta \beta^\prime}$ will involve $\gamma_0/\gamma$ when transforming back to the original coordinate. The general form of the collective Hamiltonian reads
\begin{gather}
        L(\xi,\xi^\prime)=\delta(\xi-\xi^\prime)\left[-\frac{1}{\sqrt{\gamma}}\frac{\partial}{\partial \xi}\sqrt{\gamma}\frac{1}{2M(\xi)}\frac{\partial}{\partial \xi}+V(\xi)\right], 
\end{gather}
where the collective mass and collective potential have the same forms as in Eq. (\ref{collMV}), with $\gamma_0$ replaced by $\gamma$.

\section{\label{sec:MCH-CDFT} Microscopic Collective Hamiltonian Based on Density Functional Theory}

In the previous section the collective Bohr Hamiltonian was derived starting from a microscopic GCM framework. To illustrate possible applications of the collective Hamiltonian to various low-energy structure and dynamical phenomena, selected examples that use different collective coordinates will be presented. The Hamiltonian that will be employed in the following examples is constructed using 
covariant density functional theory (CDFT, details can be found in Chapter 6) but, of course, it can equally well be built from any of a number of successful non-relativistic density functionals. In the first representative case, the quadrupole five-dimensional collective Hamiltonian (5DCH) is applied to the phenomenon of  shape coexistence in the neutron-deficient nucleus $^{76}$Kr. The spectroscopy of pear-shaped nuclei can be described using a quadrupole-octupole collective Hamiltonian. In the final example, nuclear fission dynamics is modeled by the time-evolution of a collective wave packet in the space of quadrupole and octupole axially symmetric deformations.

\subsection{\label{subsec:5DCH} The five-dimensional collective Hamiltonian}

To describe complex phenomena that originate in collective quadrupole excitations of nuclei, e.g., shape phase transitions, shape coexistence, and super-deformed bands, the simple Bohr Hamiltonian of Eq. (\ref{Hcoll}) has to be extended to a general five-dimensional collective Hamiltonian (5DCH),
\begin{align}\label{5DCH}
\hat{H} (\beta, \gamma,\theta) &=\hat{T}_{\text{vib}}+\hat{T}_{\text{rot}} +V_{\text{coll}}\nonumber\\
&=-\frac{\hbar^2}{2\sqrt{wr}} \left\{\frac{1}{\beta^4}\left[\frac{\partial}{\partial\beta}\sqrt{\frac{r}{w}}\beta^4  B_{\gamma\gamma} \frac{\partial}{\partial\beta}- \frac{\partial}{\partial\beta}\sqrt{\frac{r}{w}}\beta^3 B_{\beta\gamma}\frac{\partial}{\partial\gamma} \right]\right.\nonumber\\
& \ \ \ \left.+\frac{1}{\beta\sin{3\gamma}} \left[-\frac{\partial}{\partial\gamma} \sqrt{\frac{r}{w}}\sin{3\gamma}  B_{\beta \gamma}\frac{\partial}{\partial\beta} +\frac{1}{\beta}\frac{\partial}{\partial\gamma} \sqrt{\frac{r}{w}}\sin{3\gamma} B_{\beta \beta}\frac{\partial}{\partial\gamma} \right]\right\}\nonumber\\
&\ \ \ +\frac{1}{2}\sum_{k=1}^3{\frac{\hat{J}^2_k}{\mathcal{I}_k}}+V_{\text{coll}}.
\end{align}
where $\hat{J}_k$ denotes the components of the angular momentum in the body-fixed frame of a nucleus, and both the mass parameters $B_{\beta\beta}$, $B_{\beta\gamma}$, $B_{\gamma\gamma}$,  and the moments of inertia $\mathcal I_k$, depend on the quadrupole deformation variables $\beta$ and $\gamma$. $V_{\text{coll}}$ is the collective potential. Two additional quantities that appear in $\hat T_{\text{vib}}$, namely $r=B_1B_2B_3$ ($B_k$ are related to $\mathcal I_k$~\cite{Niksic2009}), and $w=B_{\beta\beta}B_{\gamma\gamma}-B_{\beta\gamma}^2 $, determine the volume element in the collective space. Details about the DFT-based 5DCH can be found in Refs.~\cite{Niksic2009,Libert1999,Prochniak2004}.

The 5DCH describes quadrupole vibrations, rotations, and the coupling of these collective modes.  The corresponding eigenvalue equation  is solved by expanding the eigenfunctions on a complete set of basis functions that depend on the deformation variables $\beta$ and $\gamma$, and the Euler angles \cite{Prochniak2004}. The result are the excitation energy spectrum $E^I_\alpha$ and the collective wave functions
\begin{align}
\Psi^{IM}_\alpha(\beta, \gamma, \theta)=\sum_{K\in\Delta I}\psi^I_{\alpha K}(\beta, \gamma)\phi^I_{MK}(\theta),
\end{align}
where $M$ and $K$ are the projections of angular momentum $I$ on the third axis in the laboratory and intrinsic frames, respectively, and $\alpha$ denotes additional quantum numbers. The rotational function $\phi^I_{MK}(\theta)$ is defined by Eq. (\ref{eq:phi}).
Using the collective wave functions, various observables, such as $E2$ transition probabilities, can be calculated,
\begin{align}
B(E2;\alpha I\rightarrow\alpha'I')=\frac{1}{2I+1}|\langle\alpha'I'||\hat M(E2)||\alpha I\rangle|^2 ,
\end{align}
where $\hat M (E2)$ is the electric quadrupole operator.

The next step is to calculate the collective inertia parameters $B$ and $\mathcal I_k$, and the collective potential $V_{\rm coll}$, using solutions of microscopic self-consistent deformation-constrained triaxial CDFT calculations (c.f. Chapter 6). In principle, one can use results obtained with the GCM+GOA framework. However, in practice, the resulting inertia parameters are generally too small \cite{Baranger1978,Goeke1980}. The adiabatic time-dependent
Hartree-Fock (ATDHF) theory \cite{Baranger1978} provides an alternative way to derive a classical collective Hamiltonian, and, after requantization, a Bohr Hamiltonian of the same structure is obtained but with different microscopic expressions for the inertia parameters \cite{Villars1977}. This method has the advantage that the time-odd components of the microscopic wave functions are also included and, in this sense, the full dynamics of a nuclear system. In many applications a further simplification is thus introduced in terms of cranking formulas for the inertia parameters and zero-point energy corrections \cite{Girod1979}.

The entire map of the energy surface as a function of quadrupole deformations is obtained by imposing constraints on the axial and triaxial mass quadrupole moments:
\begin{equation}\label{ES}
\langle H \rangle +\sum_{\mu=0, 2}C_{2\mu}(\langle \hat{Q}_{2\mu}\rangle-q_{2\mu})^{2},
\end{equation}
where $q_{2\mu}$ is the constrained value of the multipole moment and $C_{2\mu}$ is the corresponding stiffness constant \cite{Ring1980}. $\langle \hat{Q}_{2\mu}\rangle$ denotes the expectation value of the mass quadrupole operator:
\begin{equation}\label{QO}
\langle \hat{Q}_{20}\rangle=2z^{2}-x^{2}-y^{2}\ \ \text{and} \ \ \langle \hat{Q}_{22}\rangle=x^{2}-y^{2}.
\end{equation}

The single-nucleon wave functions, energies, and occupation factors, generated from constrained triaxial CDFT calculations, provide the microscopic input for the parameters of the collective Hamiltonian. The moments of inertia are calculated according to the Inglis-Belyaev formula \cite{Inglis1956,Belyaev1961}
\begin{equation}
\mathcal{I}_{k}=\sum_{i, j} \frac{\left(u_{i} v_{j}-v_{i} u_{j}\right)^{2}}{E_{i}+E_{j}}\left|\left\langle i\left|\hat{J}_{k}\right| j\right\rangle\right|^{2}, \quad k=1,2,3,
\end{equation}
where $k$ denotes the axis of rotation, and the summation runs over the proton and neutron quasiparticle states. $E_{i}$, $v_{i}$, and $|i\rangle$ are the quasiparticle energies, occupation probabilities, and single-nucleon wave functions, respectively. The mass parameters associated with the two quadrupole collective coordinates $q_{0}=\left\langle\hat{Q}_{20}\right\rangle$ and $q_{2}=\left\langle\hat{Q}_{22}\right\rangle$ are also calculated in the cranking approximation Ref.\cite{Girod1979}
\begin{equation}
B_{\mu \nu}\left(q_{0}, q_{2}\right)=\hbar^{2}\left[\mathcal{M}_{(1)}^{-1} \mathcal{M}_{(3)} \mathcal{M}_{(1)}^{-1}\right]_{\mu \nu},
\end{equation}
with
\begin{equation}
\mathcal{M}_{(n), \mu \nu}\left(q_{0}, q_{2}\right)=\sum_{i, j} \frac{\left\langle i\left|\hat{Q}_{2 \mu}\right| j\right\rangle\left\langle j\left|\hat{Q}_{2 v}\right| i\right\rangle}{\left(E_{i}+E_{j}\right)^{n}}\left(u_{i} v_{j}+v_{i} u_{j}\right)^{2}.
\end{equation}

The collective energy surface includes the energy of zero-point motion, which has to be subtracted. The collective zero-point energy (ZPE) corresponds to a superposition of zero-point motion of individual nucleons in the single-nucleon potential. In practice, ZPE corrections originating from the vibrational and rotational kinetic energy are considered. These are given by the following expressions \cite{Girod1979}
\begin{equation}
\Delta V_{\mathrm{vib}}\left(q_{0}, q_{2}\right)=\frac{1}{4} \operatorname{Tr}\left[\mathcal{M}_{(3)}^{-1} \mathcal{M}_{(2)}\right] ,
\end{equation}
and
\begin{equation}
\Delta V_{\mathrm{rot}}\left(q_{0}, q_{2}\right)= \Delta V_{-2-2}\left(q_{0}, q_{2}\right)+\Delta V_{-1-1}\left(q_{0}, q_{2}\right)
+\Delta V_{11}\left(q_{0}, q_{2}\right) ,
\end{equation}
with
\begin{equation}
\Delta V_{\mu \nu}\left(q_{0}, q_{2}\right)=\frac{1}{4} \frac{\mathcal{M}_{(2), \mu \nu}\left(q_{0}, q_{2}\right)}{\mathcal{M}_{(3), \mu \nu}\left(q_{0}, q_{2}\right)} .
\end{equation}
Finally, the collective potential $V_{\rm coll}$ in Eq. (\ref{5DCH}) is obtained by subtracting the ZPE corrections from the total mean-field energy:
\begin{equation}
\label{ZPEs}
V_{\text {coll }}\left(q_{0}, q_{2}\right)=E_{\text {tot }}\left(q_{0}, q_{2}\right)-\Delta V_{\text {vib }}\left(q_{0}, q_{2}\right)-\Delta V_{\text {rot }}\left(q_{0}, q_{2}\right) .
\end{equation}

\subsection{\label{subsec:SC-76Kr} Shape coexistence in $^{76}$Kr}

The observation that an atomic nucleus with a particular combination of neutron $N$, and proton $Z$ numbers, can exhibit states characterized by different shapes, i.e., shape coexistence, appears to be a unique feature of finite many-body quantum systems \cite{Heyde2011}. The low-lying states of neutron-deficient even-even krypton isotopes are of particular interest due to their rapid structural change with neutron number, and the presence of multiple shape coexistence; i.e., several $0^+$ states with different intrinsic shapes coexist at low excitation energy. The irregularities observed in the ground-state bands at low spin in $^{74, 76}$Kr were explained by shape coexistence in Ref.~\cite{Piercey1981}. As an example of the application of the quadrupole collective Hamiltonian, here the CDFT-based 5DCH model is used to analyze shape coexistence in $^{76}$Kr. A systematic study of shape evolution and shape coexistence in Kr isotopes has been reported in Ref.~\cite{Fu2013}.

\begin{figure}[htp]
	\centering
	\includegraphics[width=0.8\linewidth]{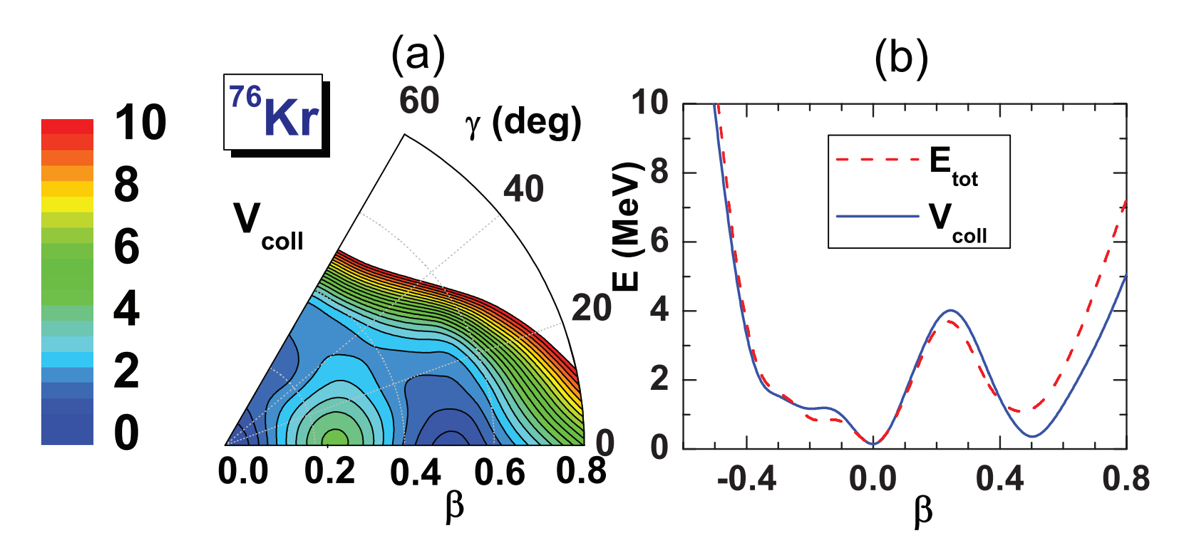}
	\caption{(a) Collective potential energy surface of $^{76}$Kr \cite{Fu2013} in the $\beta$-$\gamma$ plane calculated using the constrained triaxial CDFT with the PC-PK1 functional \cite{Zhao2010}. (b) Comparison of the total axial deformation energy and the collective potential of$^{76}$Kr. All energies are normalized to that of the spherical shape.}
	\label{fig:PES}
\end{figure}

Figure~\ref{fig:PES} displays the collective potential energy surface in the $\beta$-$\gamma$ plane for the even-even nucleus $^{76}$Kr, with the ZPEs of rotational and vibrational motion subtracted from the total deformation energy [cf. Eq.~(\ref{ZPEs})]. As shown in the panel on the right, after subtraction of the ZPEs the prolate deformed minimum becomes deeper, and the energy with respect to the spherical shape is reduced by $\approx 0.9$ MeV. This leads to a coexistence picture of competing spherical and prolate minima on the energy surface. When correlations related to restoration of broken symmetries are taken into account in the 5DCH calculation, the ground state of $^{76}$Kr is dominated by the prolate deformed configurations.

\begin{figure}
	\centering
	\includegraphics[width=1\linewidth]{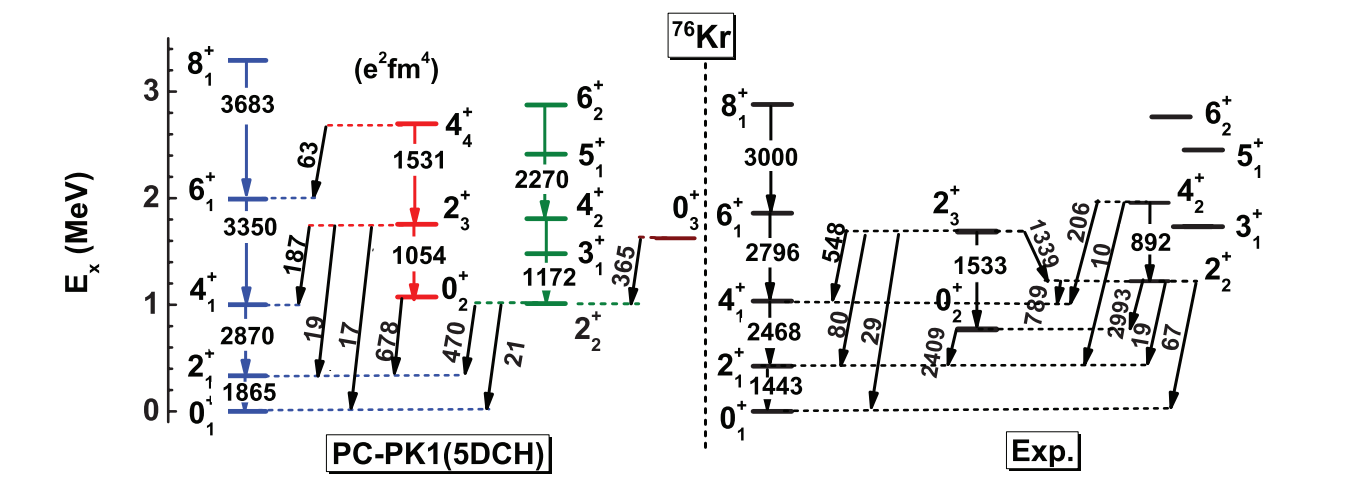}
	\caption{ (Color online) Low-spin excitation spectrum of $^{76}$Kr \cite{Fu2013}, in comparison with available data~\cite{Clement2007,NuclearData}. The $E2$ transition strengths are in units $e^2$ fm$^4$.}
	\label{fig:spectra}
\end{figure}

The 5DCH low-spin excitation spectrum of the nucleus $^{76}$Kr is compared with available data in Fig.~\ref{fig:spectra}. The main features are reproduced very well by the model calculation, in particular for the ground-state band and the low-lying $0^+_2$ state. The observed large $E2$ transition strength from the $0^+_2$ state to the $2^+_1$ state is underestimated by about a factor of four. This may imply that the mixing between the two calculated bands is too weak. In contrast, the large $B(E2;  0^+_2\to 2^+_1)$ can be reproduced by a similar 5DCH calculation that uses the Gogny D1S functional, which predicts a rather large mixing between the two $0^+$ bands,  characterized by a very large electric monopole transition rate $\rho(E0; 0^+_2\to 0^+_1)$ \cite{Clement2007}.

\begin{figure}
	\centering
	\includegraphics[width=1\linewidth]{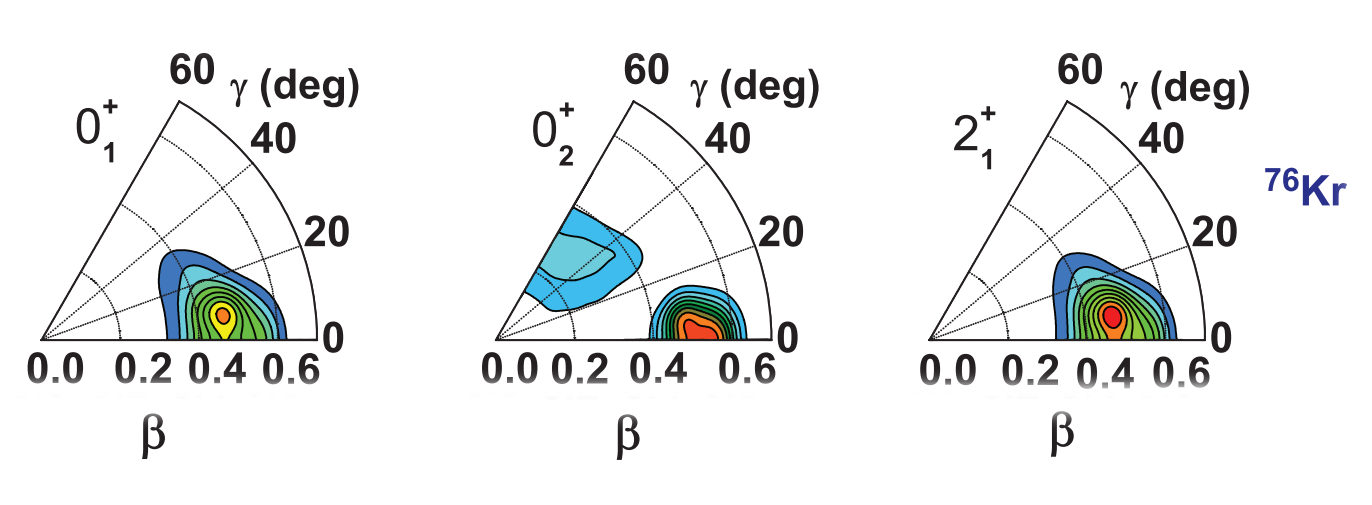}
	\caption{Distribution of the probability density $\rho_{I \alpha}(\beta,\ \gamma)$ for the first two $0^+$ states, and the $2_1^+$ state in $^{76}$Kr \cite{Fu2013}.}
	\label{fig:wave}
\end{figure}

The density distribution of the collective states, which takes the following form,
\begin{align}
    \rho_{I\alpha}(\beta,\gamma)=\sum_{K\in\Delta I}|\psi^I_{\alpha K}(\beta,\gamma)|^2\beta^3,
\end{align}
with the normalization
\begin{align}
    \int^\infty_0\beta d\beta\int^{2\pi}_0\rho_{I\alpha}(\beta,\gamma)|\sin(3\gamma)|d\gamma=1,
\end{align}
provides further insight into shape coexistence. Figure \ref{fig:wave} displays $\rho_{I\alpha}$ in the $\beta$-$\gamma$ plane for the first two $0^+$ states, and the $2^+_1$ state in $^{76}$Kr. Obviously, the dominant configurations of the ground state band correspond to the large prolate deformed shape at $(\beta\sim0.50)$. The distribution of probability density $\rho_{I \alpha}(\beta,\ \gamma)$ indicates a prolate-oblate mixed configuration for the $0^+_2$ state. From the potential energy surface shown in Fig.~\ref{fig:potential}, one notices that $^{76}$Kr is rather soft with respect to $\gamma$ deformation, and this is reflected in the structure of the $0^+_2$ state.
\vspace{-0.5cm}


\vspace{-0.5cm}
\subsection{\label{subsec-QOCH} Quadrupole-octupole collective Hamiltonian for pear-shaped nuclei}

\begin{figure}[ht]
	\begin{center}
		\includegraphics[height=0.5\textwidth]{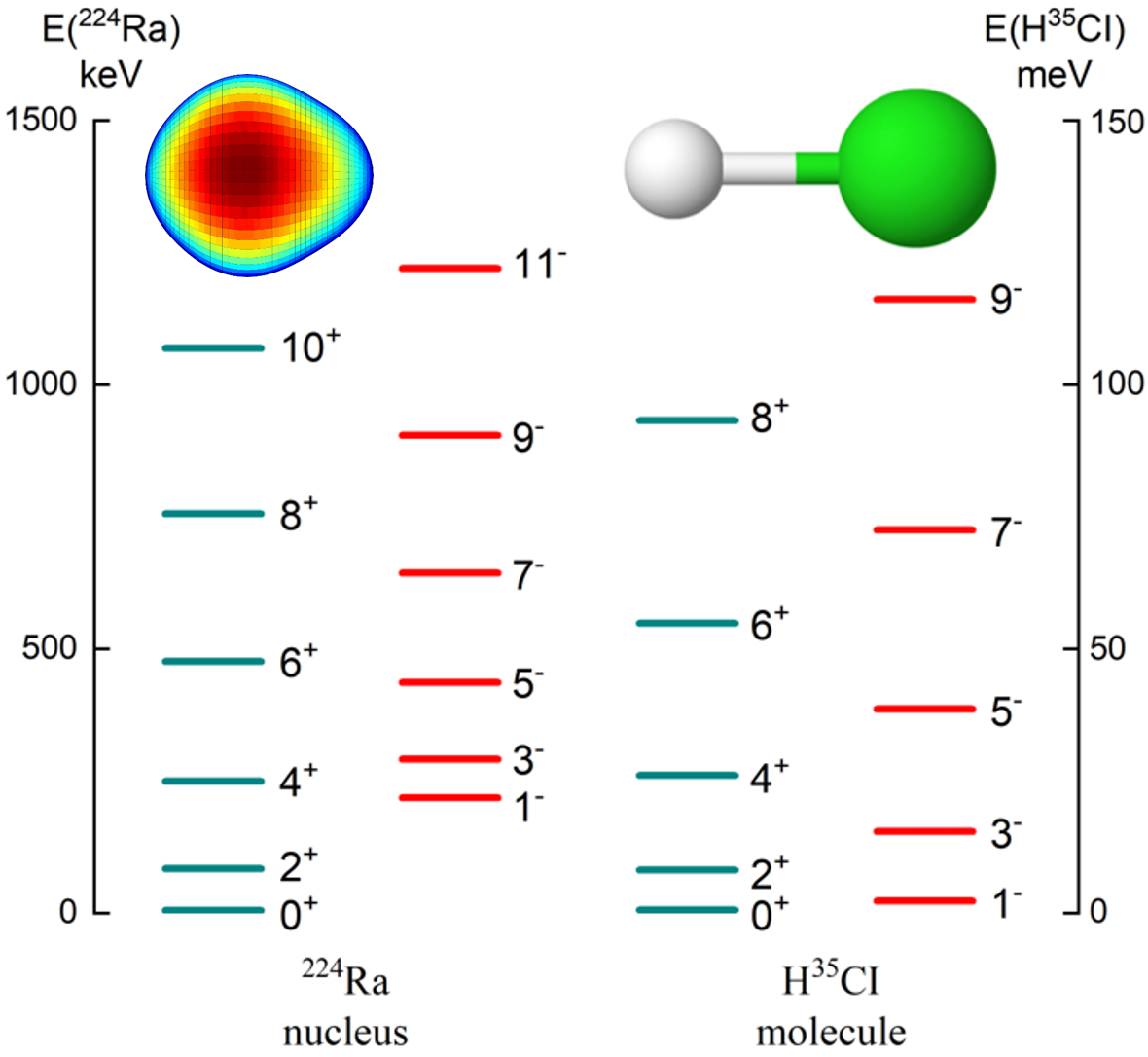}
		\caption{The low-lying rotational spectrum of $^{224}$Ra, compared with that of the H$^{35}$Cl molecule \cite{Butler1996}. Schematic shapes for the two systems are also shown.}
		\label{fig:oct-diagram}
	\end{center}
\end{figure}

Even though most deformed medium-heavy and heavy nuclei exhibit quadrupole, reflection-symmetric equilibrium shapes, there are regions in the mass table where octupole deformations (pear shapes) occur, in particular, nuclei with neutron (proton) number $N(Z)\approx 34, 56, 88$ and $134$. Pear shapes are characterized by the occurrence of low-lying negative-parity bands [c.f. Fig. (\ref{fig:oct-diagram})], as well as pronounced electric octupole transitions \cite{Gaffney2013Nature199}. The physics of octupole correlations was extensively explored several decades ago (see the review of Ref. \cite{Butler1996}), but there has also been a strong revival of interest in pear shapes more recently, as shown by a series of experimental studies \cite{Butler2019Nature,Butler2020PRL,Chishti2020Nature}.

The quadrupole-octupole collective Hamiltonian (QOCH), which can simultaneously treat the axially symmetric quadrupole and octupole vibrational and rotational excitations, is expressed in terms of two deformation parameters $\beta_2$ and $\beta_3$, and the Euler angles $\theta$ that define the orientation of the intrinsic principal axes in the laboratory frame,
\begin{equation}
\begin{split}
{{\hat{H}}_{\text{coll}}}=&-\dfrac{\hbar^2}{2\sqrt{w\mathcal{I}}}
\left[\dfrac{\partial}{\partial \beta_2}\sqrt{\dfrac{\mathcal{I}}{w}}B_{33}\dfrac{\partial}{\partial \beta_2}
-\dfrac{\partial}{\partial \beta_2}\sqrt{\dfrac{\mathcal{I}}{w}}B_{23}\dfrac{\partial}{\partial \beta_3}\right. \\
&\left.-\dfrac{\partial}{\partial \beta_3}\sqrt{\dfrac{\mathcal{I}}{w}}B_{23}\dfrac{\partial}{\partial \beta_2}
+\dfrac{\partial}{\partial \beta_3}\sqrt{\dfrac{\mathcal{I}}{w}}B_{22}\dfrac{\partial}{\partial \beta_3}\right] \\
&+\dfrac{\hat{J}^2}{2\mathcal{I}}+{{V}_{\text{coll}}}( {{\beta }_{2}}, {{\beta }_{3}} ).
\end{split}
\label{eq:QOCH}
\end{equation}
$\hat{J}$ denotes the component of angular momentum perpendicular to the symmetry axis in the body-fixed frame of a nucleus. The mass parameters $B_{22}$, $B_{23}$, and $B_{33}$, the moments of inertia $\mathcal{I}$, and collective potential $V_\text{coll}$, depend on the quadrupole and octupole deformation variables $\beta_2$ and $\beta_3$.
The additional quantity that appears in the vibrational kinetic energy, $w=\sqrt{B_{22}B_{33}-B_{23}^2}$, determines the volume element in the collective space. Just as in the case of the 5DCH model, all the collective parameters are calculated from the self-consistent solutions of constrained reflection-asymmetric CDFT, using cranking formulas \cite{Xia2017}.

\begin{figure}[htp]
	\begin{center}
		\includegraphics[height=0.5\textwidth]{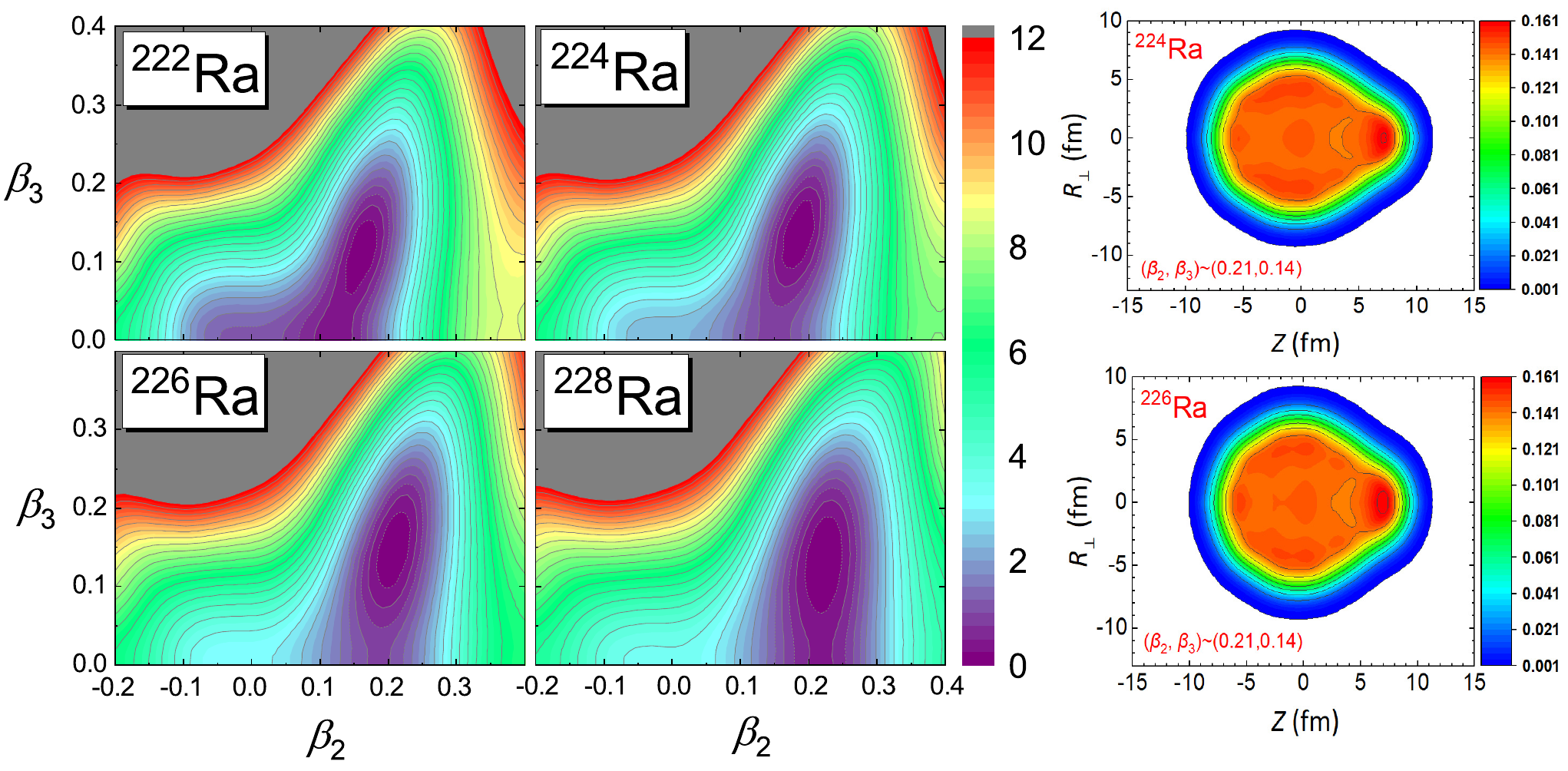}
		\caption{\label{fig:pesRa} Deformation energy surfaces of the nuclei $^{222-228}$Ra \cite{Sun2019} in the $\beta_2$-$\beta_3$ plane calculated with the CDFT, using the PC-PK1 functional \cite{Zhao2010}. For each nucleus the energies are normalized with respect to the binding energy of the global minimum. Contour lines are separated by 0.5 MeV (between neighboring solid curves) and 0.25 MeV (between neighboring dashed and solid curves), respectively. The density distributions of the global equlibrium minima of $^{224, 226}$Ra are also shown in panels on the right.}
	\end{center}
\end{figure}
As an illustrative example, the CDFT-based QOCH is used to calculate the low-energy excitation spectra and electromagnetic transitions of Ra isotopes \cite{Sun2019}. In Fig.~\ref{fig:pesRa} it is shown that already at the mean-field level the calculation predicts a very interesting structural evolution in Ra isotopic chain. Quadrupole deformation increases starting from $^{222}$Ra, and one also notices the emergence of octupole deformation with $\beta_3\sim0.11$. For $^{224, 226}$Ra the occurrence of a rather strongly marked octupole minimum is predicted. The deepest octupole minimum is calculated in $^{226}$Ra and the octupole deformation energy is $\sim0.94$ MeV. In $^{228}$Ra the deformation energy surface exhibits a softer minimum in the $\beta_3$ direction, and the octupole deformation starts to decrease.

\begin{figure}[htp]
	\begin{center}
		\includegraphics[height=0.8\textwidth]{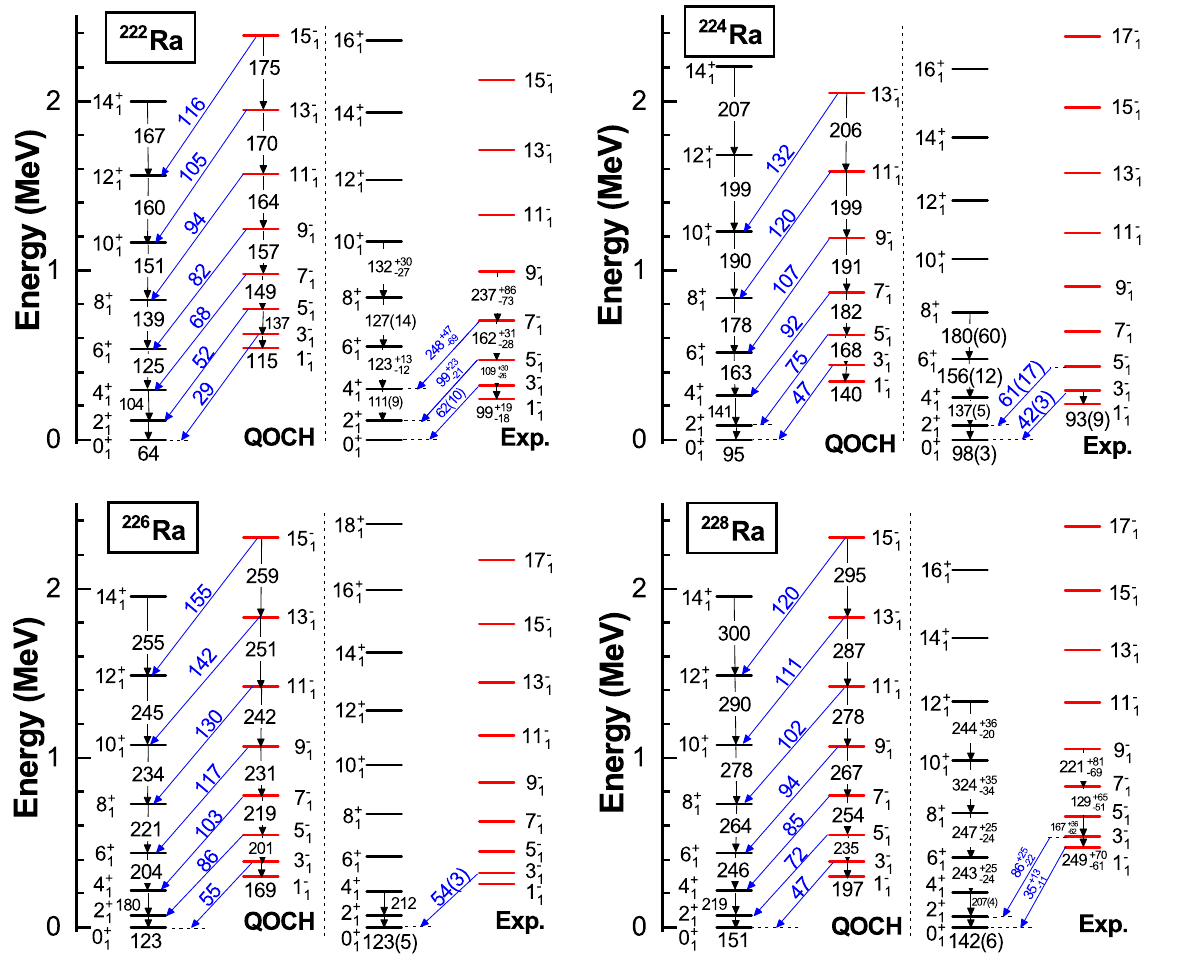}
		\caption{(Color online) The excitation spectrum, intraband $B(E2)$ (in W.u.), and interband $B(E3)$ (in W.u.) values of $^{222-228}$Ra, calculated with the QOCH based on the functional PC-PK1, and compared to experimental results \cite{NuclearData,Gaffney2013Nature199,Butler2020PRL}.}
		\label{fig:ra224spec}
	\end{center}
\end{figure}

In Fig.~\ref{fig:ra224spec} the low-energy excitation spectra of positive- and negative-parity states, the corresponding $B(E2)$ values for intraband transitions, and the interband $B(E3)$ values, calculated with the QOCH based on the PC-PK1 functional, are compared with data for the nuclei $^{222-228}$Ra \cite{Gaffney2013Nature199}. The level schemes show that the lowest negative-parity bands are located close in energy to the corresponding ground-state positive-parity bands. In fact, one notices that the lowest positive- and negative-parity bands form a single, alternating-parity band, starting with angular momentum $J \approx 5$. Except for the negative-parity band-heads in $^{222}$Ra and $^{228}$Ra that are calculated somewhat higher and lower than their corresponding experimental counterparts, respectively, the overall structure agrees very well with the available data. The calculated $E2$ and $E3$ transition rates are also in reasonable agreement with the experimental values. The shape transition is also confirmed by other characteristic collective observables, such as the $B(E3; 3^-_1\to 0^+_1)$ and $B(E2; 2^+_1\to 0^+_1)$. The $B(E2; 2^+_1\to 0^+_1)$ values increase from $^{222}$Ra to $^{228}$Ra, indicating an enhancement of quadrupole collectivity. The theoretical $B(E3; 3^-_1\to 0^+_1)$ values can be used as a measure of octupole collectivity, and exhibit a maximum in $^{226}$Ra, together with $E3$ rates for higher spin states.

%
\begin{figure}[]
	\centering
	\includegraphics[width=12cm]{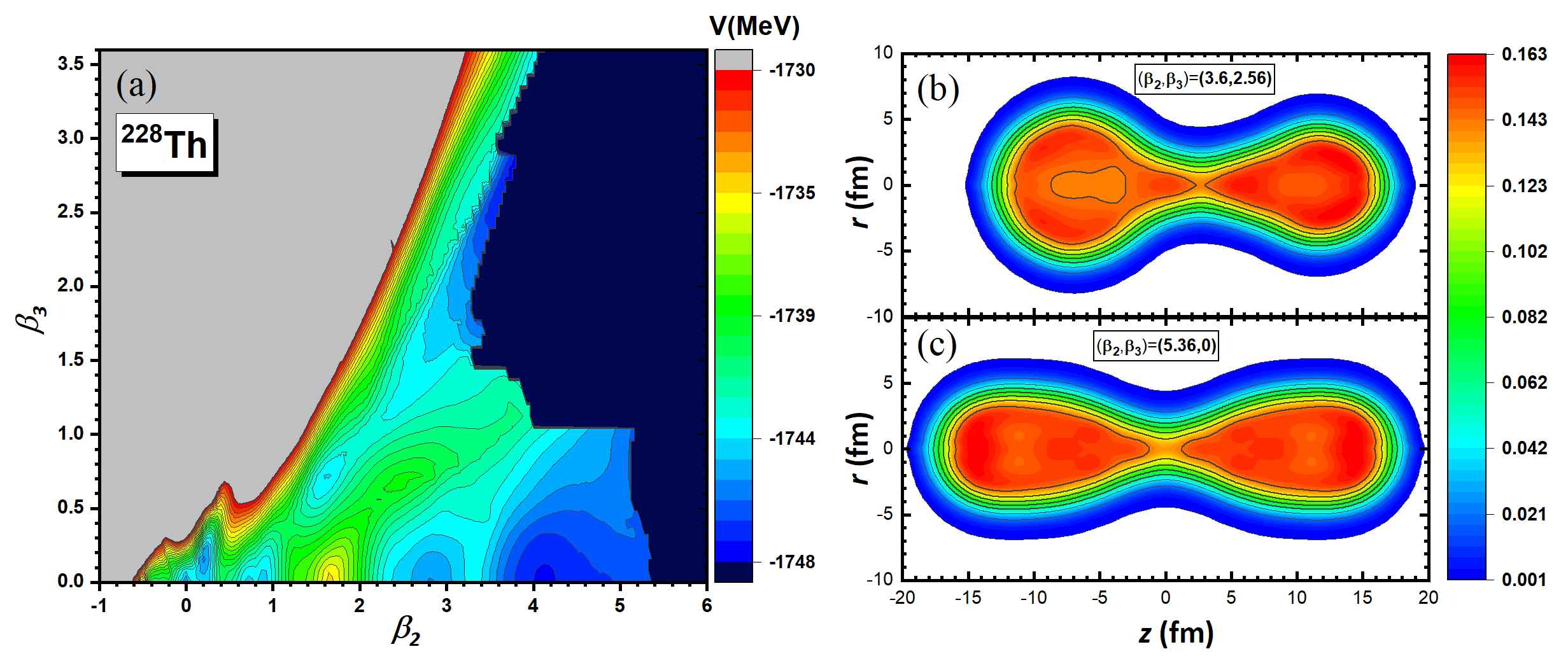}
	\caption{Potential energy surface of $~^{228}$Th in the $(\beta_2, \beta_3)$ plane, calculated using the reflection-asymmetric CDFT with the PC-PK1 functional. The nucleon density distributions at $(\beta_2,\beta_3)=(3.6, 2.56)$ and $(5.36, 0)$, are shown in the panels on the right.}
	\label{PES}
\end{figure}

\subsection{\label{TDGCM} Time-dependent GCM+GOA for nuclear fission}
As a final example, the time-dependent generator coordinate method plus Gaussian overlap approximation (TDGCM+GOA) is applied to nuclear fission dynamics \cite{Tao2017,Verriere2020,Schunck2016}. 
In the exothermic process of fission decay, an atomic nucleus splits into two or more independent fragments. A quasi-stationary initial state evolves in time through a sequence of increasingly deformed shapes. The system reaches the outer saddle point on the deformation energy surface, and continues to deform while a neck appears that eventually becomes so thin that scission occurs. Fission 
can be described as a slow adiabatic process determined by only a few collective degrees of freedom, for example the quadrupole $\beta_2$ and octupole $\beta_3$ deformation parameters. 
In the TDGCM+GOA fission dynamics is modeled by a local, time-dependent Schr\"odinger-like equation in the space of collective coordinates 
\begin{equation}
  \label{eq:TDGCM+GOA}
  i\hbar\frac{\partial g(\beta_2,\beta_3,t)}{\partial t}=\hat{H}_{\rm coll}(\beta_2,\beta_3)g(\beta_2,\beta_3,t) ,
\end{equation}
with the collective Hamiltonian
\begin{equation} \hat{H}_{\rm coll}(\beta_2,\beta_3)=-\frac{\hbar^2}{2}\sum_{kl}\frac{\partial}{\partial\beta_k}[B^{-1}(\beta_2,\beta_3)]_{kl}\frac{\partial}{\partial\beta_l}+V(\beta_2,\beta_3) .
\end{equation}
$g(\beta_2,\beta_3,t)$ is a time-dependent complex wave function in the $(\beta_2,\beta_3)$ plane. Eq.(\ref{eq:TDGCM+GOA}) describes how an atomic nucleus characterized by the collective mass $B_{kl}$ evolves in time in the collective potential $V$. As an illustration, the left panel of Fig.\ref{PES} displays the potential energy surface of $~^{228}$Th, calculated using CDFT with the PC-PK1 functional \cite{Zhao2010}. From $(\beta_2, \beta_3)\approx (0.9, 0.0)$ to $\approx (3.6, 2.56)$, an asymmetric fission valley extends on the energy surface, with two saddle points located at $(\beta_2, \beta_3)\approx (1.2, 0.35)$ and $\approx (2.1, 1.0)$. For elongations $\beta_2 > 2.0$, a symmetric valley extends up to the scission point at $\beta_2\approx 5.36$. The symmetric and asymmetric fission valleys are separated by a ridge from $(\beta_2, \beta_3)\approx (1.6, 0.0)$ to $\approx (3.8, 1.2)$.

\begin{figure}[]
	\centering
	\includegraphics[width=12cm]{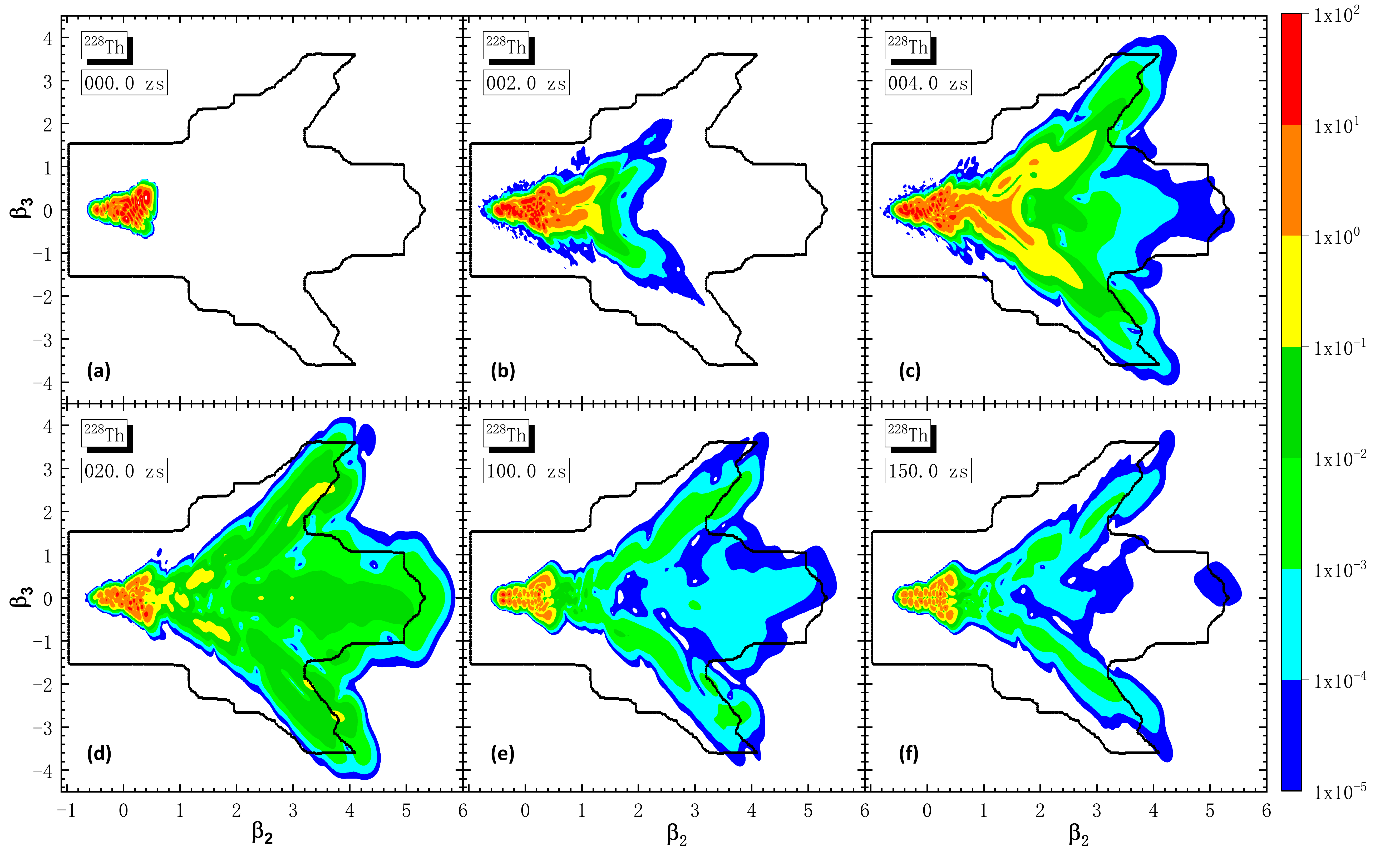}
	\caption{Time evolution of the probability density $|g|^2$ in the $(\beta_2, \beta_3)$ plane. The solid contour corresponds to the scission hypersurface.}
	\label{fig:transformation}
\end{figure}

In the panels on the right of Fig. \ref{PES}, the nucleon density distribution is plotted at $(\beta_2,\beta_3)=(3.6, 2.56)$ and $(5.36, 0)$, two point on the potential energy surface that are close to scission. The nascent fission fragments can be clearly identified and, therefore, the number of particles in the connecting neck can be defined
\begin{equation}
	Q=\int_0^{2\pi}d\varphi\int_0^\infty rdr\int_{-\infty}^{+\infty}\rho(r,\varphi,z)exp[-(z-z_n)^2]dz
\end{equation}
where $\rho$ is the nucleon density distribution in cylindrical coordinates, and $z_n$ is the point of minimum density on the z-axis. When the number of nucleons in the neck region falls below a certain value, typically few nucleons, scission occurs. Considering the entire PES, the scission points define the scission hypersurface. Given an appropriate initial state, Eq. (\ref{eq:TDGCM+GOA}) evolves the collective wave function from the inner region with a single nuclear density distribution, to the 
external region that contains the two fission fragments. 
(see the illustration in Fig. \ref{fig:transformation}).

\begin{figure}[]
	\centering
	\includegraphics[width=8cm]{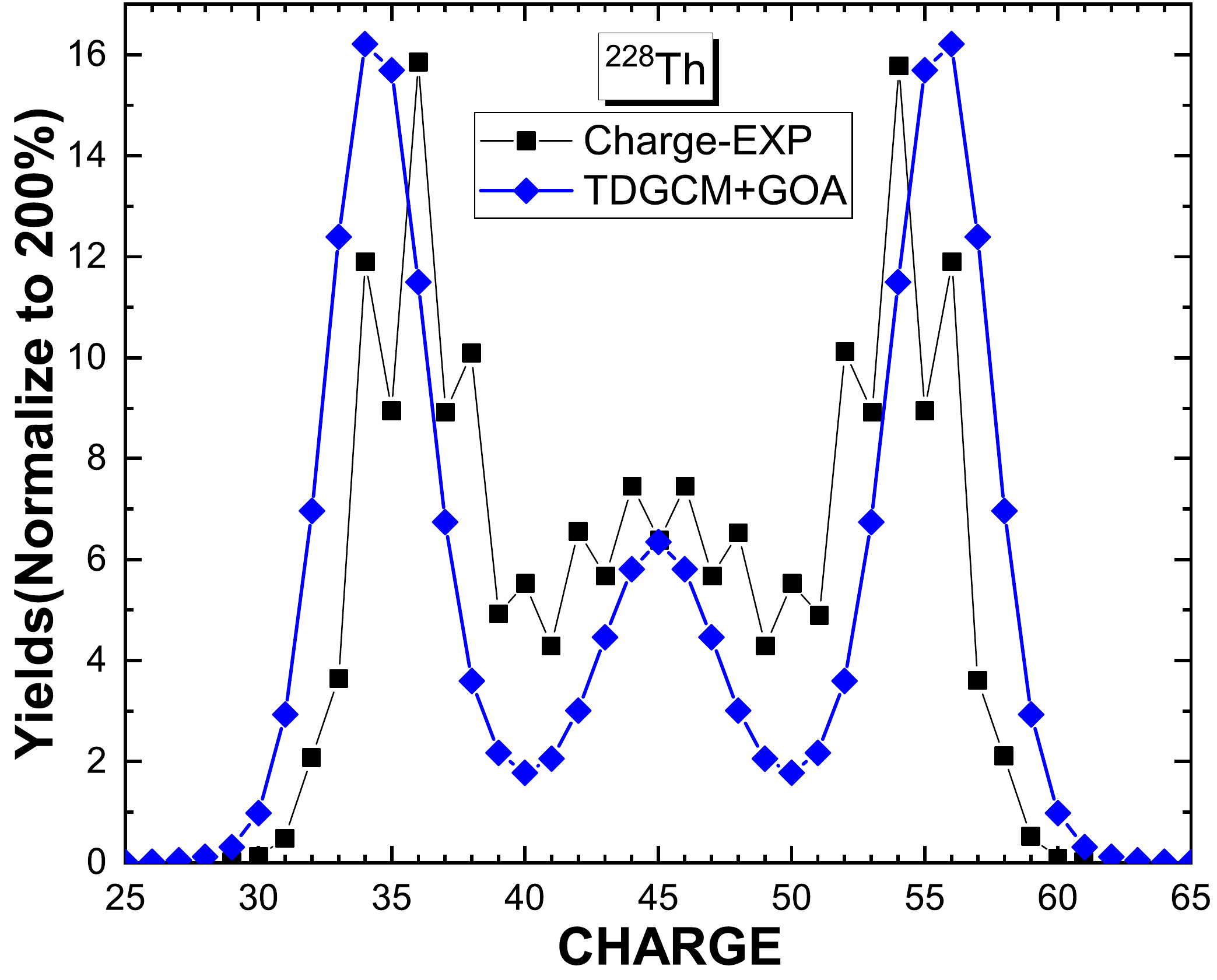}
	\caption{The charge yields (normalized to 200\%) for induced fission of $~^{228}$Th, calculated using the TDGCM+GOA, in comparison with data of photo-induced fission at an excitation energy $\approx 11$ MeV \cite{Schmidt2001}.}
	\label{yields}
\end{figure}

The flux of the probability current through the scission 
hypersurface provides a measure of the probability of observing a given pair of fragments at time $t$. The expression for the probability current $J$ in the collective space reads
\begin{equation}
	J_k(\beta_2,\beta_3,t)=\frac{\hbar}{2i}\sum_{l=2}^3 B_{kl}(\beta_2,\beta_3)[g^*(\beta_2,\beta_3,t)\frac{\partial g(\beta_2,\beta_3,t)}{\partial\beta_l}-g(\beta_2,\beta_3,t)\frac{\partial g^*(\beta_2,\beta_3,t)}{\partial\beta_l}] .
\end{equation}
The integrated flux for a given scission surface element associated with a specific pair of fragments is computed from 
\begin{equation}
	Y=\int_0^\infty dt \vec{J}\cdot\vec{ds} .
\end{equation}
By considering all points on the scission contour, the final charge or mass distribution of fission fragments is obtained. Figure \ref{yields} displays the charge yields (normalized to 200\%) for induced fission of $~^{228}$Th, calculated using the TDGCM+GOA, in comparison with data of photo-induced fission at an excitation energy $\approx 11$ MeV \cite{Schmidt2001}. The calculation reproduces the trend of the data except that, without particle number projection, the model obviously cannot describe the odd-even staggering of the experimental charge yields.


\section{\label{sec:FR} Further Reading}
This chapter has focused on the basic features of large amplitude collective motion in atomic nuclei. 
Various additional topics related to nuclear collective motion are thoroughly covered in the following textbooks:
\begin{itemize}
\item P. Ring and P. Schuck, The nuclear many-body problem, Springer-Verlag, 2004.
\item W. Greiner and J. A. Maruhn, Nuclear models, Springer-Verlag, 1996.
\item D. J. Rowe, Nuclear collective motion, World Scientific, 2010. 
\item D. J. Rowe and J. L. Wood, Fundamentals of nuclear models, World Scientific, 2010.
\item W. Younes and W. D. Loveland, An introduction to nuclear fission, Springer-Verlag, 2021.
\end{itemize}
Different aspects of quadrupole collective dynamics, and more detailed derivations of specific implementations of the collective model, are discussed in review articles:
\begin{itemize}
\item L. Pr\' ochniak, and S. G. Rohozi\' nski, Quadrupole collective states within the Bohr collective Hamiltonian, J. Phys. G: Nucl. Part. Phys. 36, 123101 (2009). 
\item		T. Nik\v si\' c, D. Vretenar, and P. Ring, Relativistic nuclear energy density functionals: Mean-field and beyond, Prog. Part. Nucl. Phys. 66, 519 (2011). 
\item		S. G. Rohozi\' nski, Gaussian overlap approximation for the quadrupole collective states, J. Phys. G: Nucl. Part. Phys 39, 095104 (2012).
\item		S. G. Rohozi\' nski, On the Gaussian overlap approximation for the collective excitations of odd nuclei, J. Phys. G: Nucl. Part. Phys. 42, 025109 (2015).  
\item		K. Matsuyanagi, M. Matsuo, T. Nakatsukasa, K. Yoshida, N. Hinohara, and K. Sato, Microscopic derivation of the quadrupole collective Hamiltonian for shape coexistence/mixing dynamics, J. Phys. G: Nucl. Part. Phys 43, 024006 (2016).
\item		J. L. Egido, State-of-the-art of beyond mean field theories with nuclear density functionals, Phys. Scr. 91, 073003 (2016).
\end{itemize}

%
%
%
%

\end{document}